\documentclass[aps,pre,onecolumn,amssymb,showpacs]{revtex4}
\usepackage{amsmath}
\usepackage{amssymb}
\usepackage{amsfonts}
\usepackage{bm}
\usepackage{graphicx}

\newcommand{\sembrack}[1]{[\![#1]\!]}

\newcounter{Cequ}

\newcommand{\nv}{{\bf {\hat n}}}
\newcommand{\bseq}{\begin{subequations}}
\newcommand{\eseq}{\end{subequations}}
\newcommand{\bal}{\begin{align}}
\newcommand{\ealig}{\end{align}}
\newcommand{\be}{\begin{equation}}
\newcommand{\ee}{\end{equation}}
\newcommand{\bea}{\begin{eqnarray}}
\newcommand{\beal}{\begin{align}}
\newcommand{\eali}{\end{align}}
\newcommand{\eea}{\end{eqnarray}}
\newcommand{\bsub}{\begin{subequations}}
\newcommand{\esub}{\end{subequations}}
\newcommand{\grad}{\bm \nabla}
\newcommand{\rot}{\text{rot}}

\newcommand{\LB}[1]{\label{#1}}
\newcommand{\scal}[2]{(\bm {#1} \cdot \bm {#2})}
\newcommand{\im}{i\,}

\newcommand{\deln}{\delta\hspace{-0.031cm}\bm n}
\newcommand{\delns}{\delta\hspace{-0.031cm}n}
\newcommand{\id}{{\normalfont\hbox{1\kern-0.15em \vrule width .8pt depth-.5pt}}}

\begin{document}
\title{Optical analysis of spatially periodic patterns in nematic liquid crystals:
diffraction and shadowgraphy}
\author{Werner Pesch}
\email[]{werner.pesch@uni-bayreuth.de}
\author{Alexei Krekhov}
\affiliation{Physikalisches Institut, Universit\"at Bayreuth,
95440 Bayreuth, Germany}
\date{\today}
\begin{abstract}
Optical methods are most convenient to analyze spatially periodic patterns with wavevector $\bm q$  in a  thin layer of a nematic liquid crystal.
In the standard experimental setup a beam of parallel light with a 'short' wavelength $\lambda \ll 2 \pi/q$  passes the nematic layer.
Recording the transmitted light the patterns are either directly visualized by shadowgraphy or characterized more indirectly by the diffraction fringes due to the optical grating effects of the pattern.   
In this work we present a systematic short-wavelength analysis of these methods for the commonly used planar orientation of the optical axis of liquid crystal at the confining  surfaces.
Our approach covers general 3D experimental geometries with respect to the relative orientation of $\bm q$ and of the wavevector $\bm k$ of the incident light.
In particular the importance of phase grating effects is emphasized, which are not accessible in a pure geometric optics approach.
Finally, as a byproduct we present also an optical analysis of convection rolls in Rayleigh-B\'enard convection, where the refraction index of the fluid is isotropic in contrast to its uniaxial symmetry in nematic liquid crystals.
Our analysis is in excellent agreement with an earlier physical optics approach  by Trainoff and Cannell [Physics of Fluids {\bf 14}, 1340 (2002)], which is restricted to a 2D geometry and technically much more demanding.
\end{abstract}
\maketitle

\section{Introduction}
\label{sec:intro}
Increasing an external stress on a homogeneous fluid layer leads typically to a spontaneous generation  of spatially periodic structures (patterns) in the plane of the layer \cite{CrossHo}. The structures are characterized by a wavevector $\bm q$ and a certain amplitude, which depends on the amount of stress.  Corresponding periodic structures 
are induced  into  the refraction index of the fluid layer, which thus 
acts as an optical grating when illuminated for instance with a parallel  light beam with wavelength $\lambda \ll 2 \pi /q$. The  transmitted light may be analyzed  in terms  of the arising diffraction 
fringes. They   give  insight into the intensity  of the Fourier modes  representing the periodicity of the planforms and more indirectly into the mechanism driving the pattern forming instabilities. Alternatively,   shadowgraphy is often applied to visualize directly the periodically  distorted fluid layer.
 An important paradigm is Rayleigh-B\'enard
convection (RBC) driven by a temperature gradient \cite{annual}, where the common  array of convection rolls is  mapped to a sequence of  black and white stripes in shadowgraphy \cite{bruyn96}.

 In this paper, however, we refer mainly to patterns in nematic liquid crystals, which are anisotropic uniaxial fluids.
The preferred direction (roughly speaking the mean orientation of the non-spherical molecules in the nematic phase) is described  by the director
$\nv$ with $\nv^2 =1$, which also determines the local optical axis.  
The nematic layer of thickness $d$ and large lateral extension is confined between two coplanar glass plates (parallel to the 
$xy$-plane).  They are specially  treated  to enforce  a fixed  director orientation $\nv = \bm n_0$ at the two surfaces at $z =0, d$. We consider exclusively the so called {\it planar} director configuration where $\bm n_0$    points  
along the $x$-axis (parallel to the unit vector $\bm{\hat x}$). 
 In the basic state a uniform orientation of $\nv = \bm n_0$ is then  induced throughout the whole nematic layer, corresponding to a minimum of the orientational elastic free energy.
A thoroughly studied pattern forming instability of the basic state is the electro convection instability (EC), where the glass plates  are coated in addition with  thin transparent electrodes to apply an ac voltage $U = U_0 \cos (\Omega t)$ across the layer. For a voltage   amplitude  $U_0$ above a certain threshold $U_c$  stripe pattern appear in EC \cite{ehcrev}.   But also the application of magnetic fields, temperature gradients or shearing  the  nematic layer leads easily to patterns (see for instance \cite{AbuLk}). 
Revealing the character and the mechanisms of the instabilities requires to relate the director distortions and the corresponding optical signals.

The theory of  shadowgraphy has been developed in several steps over the years. 
At first geometric (ray) optics has been applied to RBC in \cite{jenkins88}. There a simple model was proposed to describe the deflection of the incoming light rays towards the optically 
denser  (cold) regions of the RBC convection patterns. 
The model makes use of the fact, that the scale of the spatial variations of the refraction index in the fluid layer (of the order of $0.5\, \mathrm{cm}$ in typical RBC experiments)    is much larger than the light wavelength $\lambda \approx  0.6 \mu\mathrm{m}$ used. 
Since the effects of diffraction are neglected, this theory predicts     
divergent intensities (caustics) in pictures recorded at a certain level $z = z_F$ above the cell. 
 In \cite{jenkins88} $z_F$ is treated as an adjustable parameter and the theoretical pictures calculated for some  distance $z < z_F$ above the cell, look very similar  to the experimental ones.
The analysis in RBC on the basis of ray optics has been considerably refined  
by Rehberg and coworkers \cite{rasenat89,rehberg89}  by calculating the ray paths
using Fermat's principle.  
In addition they considered also shadowgraphy for optically anisotropic 
nematics on the example of EC pattern. 
Here the scale of the spatial variations of the refraction index is typically governed  by the thickness of the nematic layer (typically $10 \mu\mathrm{m} \lesssim d \lesssim 100 \mu\mathrm{m}$) and thus still considerably larger than $\lambda \approx 0.6 \mu\mathrm{m}$ of the light used.  Roughly
speaking, in \cite{rasenat89,rehberg89} a theoretical description  of $z_F$ 
as function of the pattern amplitudes and the two nematic refraction indices is achieved. Focusing in experiments  to positions $z < z_F$ above the cell, the contrast between minimal and maximal amplitudes  in shadowgraph pictures seems to agree quite well with the 
theoretical predictions.

The most comprehensive theoretical treatment of shadowgraphy so far has been presented by Trainoff and Cannell \cite{cannell}, where some problematic aspects of the prior analyses 
become evident \cite{problem}.    
In geometric optics only the amplitude grating  effects of the  patterns but not their  phase grating contributions are taken into account.  They determine for instance the weight of higher harmonics with wavevector $2 \bm q$ in shadowgraph pictures and are essential for 
the interpretation of  patterns driven by thermal fluctuations below onset of convection (see, e.g., \cite{wu95,cannell}).  
Furthermore, geometric optics is not used outside the fluid layer as in \cite{rasenat89,rehberg89} and the propagation and interference  of the light waves when leaving the fluid layer are treated rigorously. As a result, the electromagnetic patterns with maximal contrast  above the fluid layer, re-occur  at regular distances along the $z$-axis perpendicular to the layer plane. The corresponding period is given as $8 \pi^2 /(q^2 \lambda)$ and does not depend on the pattern 
amplitudes. This is  in distinct contrast to the caustics in geometric optics. Here the position of shadowgraph pictures with maximal contrast above the layer is determined by $z_F$, which even diverges when the pattern amplitude  approaches zero. 
It is noteworthy that  the periodicity of the electromagnetic field along the $z$-direction  has 
already  been described almost  two hundred  years ago by Talbot \cite{talbot}, who studied light
incident on a periodic diffraction grating.  The intriguing transition from the periodic sequence of the finite-intensity patterns along the $z-$axis   to caustics in the limit 
$\lambda  \rightarrow    0$ has been discussed recently in \cite{berry}.

In \cite{cannell} shadowgraphy in RBC has  been discussed in detail, while EC is only briefly touched. A more detailed recent investigation  \cite{stann} on EC patterns is on one hand devoted to the theoretical description of the diffraction spots (on the basis of phase grating). The main theoretical predictions are consistent  with corresponding experiments.   This paper contains also references to earlier  investigations  of the problem.
In particular  in \cite{Zengin1} the importance of phase grating had been already  stressed.

The {\it physical optics} approach in \cite{cannell,stann} is, however, quite complicated to use since amplitude- and phase grating effects are still treated separately. Furthermore the analysis  is restricted to  a special, though often utilized, 2D geometry, where $\nv$ lies in plane spanned by the wavevector $\bm q$ and the polarization of the plane wave transversing the nematic layer. 
The goal of the  present work 
 is a systematic, more transparent  analysis  of  the problem on the basis  
of the common short-wavelength approximation 
in theoretical optics \cite{Born}. The starting point  is  a  systematic  expansion of the solutions of the Maxwell equations in terms of the small parameter $\lambda q/(2 \pi)$.  We have benefited  considerably from an analysis   of light propagation in 
nematic fluid layer presented in \cite{Pana}. As a result,  for the first time diffraction and shadowgraphy for patterns in nematics are thus described  quasi-analytically for quite general 
configurations of the incident light wave (polarization, wavevector) and of the orientation of the director. 
In the case of dissipative convection rolls as in EC the presence of fluid flow plays an essential role. The wavevector $\bm q$, characterizing the pattern periodicity,  is then often parallel to the basic planar director configuration  $\bm n_0 \equiv \bm{\hat x}$  or includes only a small angle with $\bm n_0$.  Alternatively one finds so called equilibrium transitions when 
the minima  of orientational free energy exchange in the presence of electric or magnetic fields, where $\bm q \perp \bm n_0$. 
Interesting
examples are the flexoelectric  stripe patterns \cite{flexo} driven electrically or the splay-twist Freedericksz pattern \cite{Meyer} driven either by electric or magnetic fields.

The paper is organized as follows:   In Sec.~\ref {sec:general} we first review briefly the salient features of the pattern forming instabilities in nematics. Furthermore we sketch 
the theoretical background of light propagation in nematics, also in order to fix 
our notation. Section~\ref{sec:shortwave}  deals with the short-wavelength approximation
of the Maxwell equations which is appropriate to describe optics in media with a slowly varying refraction index in 
space.  Based on this approximation we re-derive and confirm in Sec.~\ref{sec:short2d} the results of \cite{cannell} in their  2D geometry.
The following two sections contain the main results of the paper, on which one might concentrate on a first reading.  
In Sec.~\ref{sec:3dpat} we discuss  the optics of three dimensional patterns. The treatment covers in particular the case  where $\bm q \perp \bm n_0$, which cannot be described by pure geometric optics.  
Section \ref{sec:discuss} is devoted to  a general discussion and summary  of our results.  The paper concludes with some final remarks also on  future perspectives in Sec.~\ref{sec:conclus}.
In several appendices, we provide  details  of our  calculations. In particular, Appendix~\ref{sec:scalopt} is devoted to our complementary optical analysis of  roll patterns in RBC.

\section{General theoretical background}
\label{sec:general}
In this section some well known basic facts on pattern forming  
instabilities in nematics are briefly summarized.  Furthermore, since the patterns  are analyzed by optical methods,  we remind on the standard description
of light propagation in uniaxial materials mainly in order to fix the notation.

\subsection{Pattern forming instabilities in nematics}
\label{subsec:pattern}

At first we address briefly some main features  of the typical  stripe patterns in nematic layers in the planar configuration defined in the introduction. 
For a sufficiently strong external stress the basic state is destabilized. The director develops a distortion $\Delta\hspace{-0.032cm}\bm n(x,y,z)$ of $\bm n_0$ for $0< z < d$ which is periodic in the $xy$-plane. The periodicity is characterized by a critical wavevector $ \bm q$, i.e. $ \Delta\hspace{-0.032cm}\bm n$ can be represented as a Fourier series
in terms of $e^{\im \bm q \cdot \bm x}$ with $\bm x =(x,y)$. 
At the confining plates, however,  the director orientation remains  
fixed (strong anchoring): $\Delta\hspace{-0.032cm}\bm n(\bm x,z=0,d) = 0$. 
In this paper
we will concentrate on the weakly nonlinear regime just above the onset of the pattern forming instability,  where the director distortion  $\Delta\hspace{-0.032cm}\bm n(\bm x,z)$ is small. 
It is convenient to introduce  the following decomposition of $\Delta\hspace{-0.031cm}\bm n$:
\be
\label{eq:Delncomp}
\Delta\hspace{-0.032cm}\bm n(\bm x,z) =-\delta n_x( \bm x,z) \,\bm{\hat  x} + \theta_m \deln  (\bm x, z)\;\text{with} \; \deln = (0,\delns_y, \delns_z) \;\text{and}\; \deln\cdot \bm{\hat x} = 0.
\ee
Exploiting the  normalization condition  $\nv^2 =1$ with $\nv = \bm n_0 + \Delta\hspace{-0.032cm}\bm n(x,y,z)$ determines $\delta n_x$ in terms of $\deln$ :
\be
\label{eq:nwink}
\nv^2  = [(1 - \delta n_x)\bm{\hat x} + \theta_m \deln]^2 = 1 \;\text{thus} \;  \delta n_x   =  \frac{1}{2} \theta_m^2 (\deln)^2 + O(\theta_m^4). 
\ee
The amplitude $\theta_m$ measures the maximal local tilt angle of $\nv$ with respect to $xy$-plane,  i.e. the out-of-plane distortion of $\bm n_0$. 
It has turned out in many cases that the director field is surprisingly well described  already within an  ``one-mode'' approximation of the following form: 
\be 
\label{eq:ansatz_n}
\nv(\bm x,z) = (1-\frac{1}{2} \theta^2_m (\deln)^2 ,0,0) + \theta_m  \deln \;, \;\;
\deln  = \vartheta(z) (0, a_y  \sin(\bm q \cdot \bm x), \cos(\bm q \cdot \bm x)) 
\ee  
with $\vartheta(z) = \sin(\pi z/d)$ to fulfill the boundary condition $\nv(\bm x,z=0,d) = \bm n_0$.  
The amplitude $a_y$ of the twist distortion ($\delta n_y)$ of the director is small for 
 dissipative patterns like in EC and of the order one for the equilibrium patterns 
like the flexoelectric ones. 
The standard optical analysis reflects  thus directly  the wavevector $\bm q$,
the amplitudes $\theta_m$, $a_y$ and two refraction indices 
$n_e$ and $n_o$ of uniaxial materials (see also following 
section \ref{subsec:light}).	 

For convenience we choose  the well studied EC as a representative example to spell some details, but our analysis makes no particular use of this special case.  
The experimental setup consists of  an extended nematic fluid layer of small thickness $d \lesssim 100 \mu m$ oriented 
parallel to the $xy$-plane, where the lateral extension are much larger than $d$.   
The amplitude coefficients $\theta_m$ and $a_y$  are in general periodic functions in time governed by the ac-frequency (up to $10^4$ Hz in some experiments) of the applied voltage.  In many cases the coefficients are already well described by their  temporal average and thus considered to be constant.  In any case the time dependence of the amplitude can be treated in an adiabatic 
approximation for optical methods, since the frequency $\omega$ of the monochromatic light waves
to probe the patterns is by many orders of magnitude larger than the ac frequency $\Omega$.
Convection sets in at a critical amplitude $U_0 = U_c$ (typically some volts), where $\bm q$ is determined by the critical wavevector $\bm q_c$   with $|\bm q_c|$ of the order of $\pi/d$.   
These  critical data as well as the detailed director configuration depend on the material parameters  of the specific nematic material, the cell thickness  and on the ac-frequency $\Omega$. They are available  from a linear stability analysis of the spatially homogeneous basic state,  which has been extensively carried through  in the last two decades (see, e.g., \cite{BoZi,BuEb}). 
The EC instability,  as most pattern forming instabilities in nematics,  is supercritical, which means that for $U_0 \gtrsim U_c$ the amplitude $\theta_m$ grows continuously with the reduced control parameter $\epsilon \equiv (U_0^2 -U_c^2)/U_c^2$  like $\sqrt{\epsilon}$.  For the nematic material MBBA, which has been used 
in the majority of EC experiments, the proportionality factor is about $80^\circ$ \cite{BoZi}, such that $\epsilon = 0.1$ corresponds to $\theta_m$ about $25^\circ$.  

To analyze the pattern forming instabilities driven by other external stresses alluded to above, one has to follow the same strategy as in EC. At first the critical stress value like $U_c$ in EC
and the critical wavevector $\bm q_c$ together with the distorted director configuration are  determined in the framework of a linear stability analysis. Then a weakly nonlinear analysis has to be used to calculate the  amplitude  of the director 
distortion as function of the  stress parameter. 
It should be realized, that the range of validity of a weakly nonlinear analysis is in general confined to small 
amplitudes $\theta_m$.  Increasing the external stress and thus $\theta_m$ leads to secondary instabilities (see, e.g., \cite{Braun}, \cite{RUD}). The resulting patterns are typically  characterized  by the appearance of  spatial periodicities with wavevectors not parallel to $\bm q_c$.  That the growth of the out-of-plane distortion of $\propto \theta_m$  with increasing $\epsilon$ must be limited is obvious, since for instance the stabilizing orientational elastic energy grows.  For the nematic material MBBA, which has been used 
in the majority of EC experiments,  the secondary instabilities set in for  angles $\theta_m$ between  
$20^\circ$ and $30^\circ$. The often complex spatio-temporal pattern developing due to such secondary instabilities are outside the scope of this paper.

\subsection{Light propagation in nematics}
\label{subsec:light}
In the following we review briefly the propagation of monochromatic light waves with circular frequency $\omega$  in a   
nematic liquid crystal, which has uniaxial optical symmetry.
In general we follow closely the notations of Born and Wolf \cite{Born}.
The starting point are the general Maxwell equations (in cgs units) where 
the factor $e^{-i \omega t}$ has been split off: 
\bseq \label{eq:maxall}
\bal
&\nabla \times  \bm H   = - i k_0 \, \bm \varepsilon \cdot \bm E, \; &\nabla \times  \bm E = i k_0 \, \bm B, \label{eq:maxa}
 \\
&\nabla \cdot \bm D = 0 , \; &\nabla \cdot \bm B = 0,\label{eq:maxb}
\end{align}
\eseq
with $ k_0 =\omega/c$ and  $c$ the vacuum speed of light. All fields depend in general on $\bm r = (\bm x,z)$.
The constitutive equations which connect the dielectric displacement
$\bm D$ to the electric field $\bm E$ and the magnetic induction $\bm B$ to the magnetic 
field $\bm H$ are given as:
\be \label{eq:const}
\bm D = \bm \varepsilon \cdot  \bm E,\; \bm B = \mu \bm H.
\ee
The magnetic permeability $\mu$ is a scalar which can be safely put to one for our materials. 
The optical dielectric tensor $\bm \varepsilon$ of the uniaxial nematics is given as:
\be \label{eq:diel}
\bm \varepsilon = {\epsilon_\perp} \bm{\mathrm{I}} + ({\epsilon_\parallel}
-{\epsilon_{\perp}}) \nv \otimes \nv.
\ee
 Here we have introduced the standard definition of  the tensor (dyadic)   product  $\bm a  \otimes \bm b$ of two vectors $\bm a, \bm b$ with the 
components $a_i b_j, \;  i,j =x,y,z$; $\bm{\mathrm{I}}$ denotes the unit matrix.  
The   frequency dependent dielectric constants ${\epsilon_\perp}, {\epsilon_\parallel}$ are taken at the frequency $\omega$. They are considered
to be real, i.e. light absorption plays no role.    The case of optically isotropic materials like glass or air  are characterized by ${\epsilon_\perp} = {\epsilon_\parallel} \equiv  \bar\epsilon$.

After eliminating $\bm B$ from Eq.~(\ref{eq:maxa})  we arrive at:
\be \label{eq:el}
\text {rot rot} \bm E =  \Delta \bm E - \nabla (\nabla \cdot \bm E) =  k_0^2 \mu \,\bm \varepsilon \cdot \bm E.
\ee
The corresponding elimination of the electric field $\bm E$ is possible as well, which 
leads to:
\be \label{eq:mag}
\text {rot} [\bm \varepsilon^{-1} \text{rot} \bm B]= k_0^2 \mu \bm B.
\ee
\vspace*{0.2cm}

\centerline {\underline{\bf Plane-wave solutions}}
\vspace*{0.5cm}
In the case of constant $\nv$ the Maxwell equations (\ref{eq:maxall}) allow for plane wave solutions  (characterized by a wavevector  $\bm k = k_0 \bm k'$), which will play 
an important role in the following.
For instance the electric field $\bm E$ has the representation:
\be \label{eq:elf}
\bm E = \bm E_0 \exp [i k_0 \bm k'  \cdot \bm r]
\; \text{with} \; \bm r = (\bm x,z) \; \text{and} \; k_0 =  \frac{\omega}{c}.
\ee 
The amplitude $\bm E_0$ is a constant vector. 
Note that the use  of       
 non-dimensionalized wavevectors (indicated by a dash) like  $\bm k' = \bm k/{k_0}$ in   Eq.~(\ref{eq:elf})  turns out to be very convenient in the following. For instance the  {\it refraction index} $n_{rf}$ fulfills 
relation $(\bm k')^2 = (n_{rf})^2$.

Representing the other fields in analogy to  Eq.~(\ref{eq:elf}) with amplitudes  $\bm D_0,
\bm B_0, \bm H_0$ 
we arrive from Eqs.~(\ref{eq:maxall}) at:
\bseq \label{eq:maxs}
\bal
 &\bm k' \times  \bm H_0  = - \bm D_0 = \bm \varepsilon \cdot \bm E_0, \;  &\bm k' \times  \bm E_0 = \mu \bm H_0, \\ &\bm k' \cdot \bm D_0 = 0 , \; &\bm k' \cdot \bm B_0 = 0.
\end{align}
\eseq
According to  Eqs.~(\ref{eq:maxs}) the vectors $\bm k', \bm D_0, \bm H_0$ are pairwise orthogonal.   Furthermore $\bm E_0$ is orthogonal to $\bm H_0$, i.e. lies in the plane of $\bm k', \bm D_0$.
It is easy to see that Eq.~(\ref{eq:el}) reduces for the plane wave $\bm E$ used in Eq.~(\ref{eq:elf})   
to:
\be \label{eq:ME}
\left [(\bm k')^2 -\bm k' \otimes \bm k' - \mu (\epsilon_\perp + (\, \epsilon_\parallel -\epsilon_\perp) \nv \otimes \nv \, ) \, \right] \cdot \bm E_0 \equiv \bm M[\bm k'] \cdot \bm E_0 = 0.
\ee

For definiteness  we concentrate in the following on the typical (planar) experimental
setup of a nematic layer parallel to the $xy$-plane with a  constant $\nv = \bm{\hat x}$ sandwiched 
by the coplanar glass plates. 
Let a monochromatic plane wave enter this configuration from  air with the wavevector
$\bm k' = {\bm k'}_\parallel + k'_z \bm{\hat e_z}, k'_z >0$. At each interface the wave is
reflected ($k'_z$ reverses sign) and diffracted, i.e. it propagates above the interface with a modified  $k'_z$. According to Snellius's law the in-plane components of the wavevector $\bm k'$ in the various material layers are equal.  The $z$-components of $\bm k'$ are determined by the respective 
refraction indices, $n_{rf}$, i.e.  by the condition  $(\bm k')^2 =  ({\bm k'}_\parallel)^2 + (k'_z)^2 = n_{rf}^2$. 

In isotropic media where $\epsilon_{\parallel} = \epsilon_{\perp} = \bar \epsilon$ we obtain from Eqs.~(\ref{eq:ME}) the well known relation
$n_{rf} = \sqrt{\bar \epsilon \mu}$. 
Since $\bar\epsilon \bm k' \cdot \bm E_0 =0$ according to Eq.~(\ref{eq:maxall}) the orientation of the electric field  vector $\bm E_0$ (the polarization) has to be perpendicular to $\bm k'$
but is otherwise undetermined. According to Eqs.~(\ref{eq:maxall}) the magnetic 
field vector, which is also perpendicular to $\bm k'$,  has to be perpendicular 
to $\bm E_0$. For finite $\bm k'_\parallel$ we speak of TM  (transverse magnetic) waves in our geometry when the 
electric  field is in the incidence plane spanned by $\bm k', \bm{\hat z}$, i.e. the magnetic field is perpendicular to that plane.  Alternatively, for the TE (transverse electric) waves
the directions of the electric and magnetic fields are interchanged. For  $\bm k'_\parallel =0$  the TM waves are defined as polarized along the $\bm{\hat x}$ axis and vice versa. Incoming waves with arbitrary polarization obviously correspond to a certain linear superposition of TM and TE waves.

In  uniaxial nematic layers   $n_{rf}$ depends in general on the {\it ordinary} refraction index, $n_o =  \sqrt{\epsilon_\perp \mu}$,   and the {\it extraordinary} one,  $n_e = \sqrt{\epsilon_\parallel \mu}$.   
In addition  the angle between $\nv$ and $\bm k'$ comes into play.
In view of the condition $\bm k'\cdot \bm D_0 =  \bm k'\cdot (\bm \varepsilon \cdot \bm E_0) =0$ (see Eq.~(\ref{eq:maxs}))     
the linear equation $\bm M\cdot  \bm E_0 = 0$ (see Eq.~\ref{eq:ME})) allows for the two linear independent solutions $\bm E_0= \bm o_0[\bm k_o']$ ({\it ordinary}) and 
  $\bm E_0 = \bm e_0[\bm k_e']$ ({\it extraordinary}) which are represented as the following unit vectors:
\be \label{eq:veco}
 \bm o_0[\bm k_o'] = \frac{\bm k_o' \times  \nv }{|\bm k_o' \times  \nv|} \;, \;\; 
\bm e_0[\bm k_e'] = \frac{   \epsilon_\perp \nv -\bm k_e' (\bm k_e' \cdot \nv) }{|\epsilon_\perp \nv -\bm k_e' (\bm k_e' \cdot \nv )|}.
\ee 
The parallel components of the wavevectors $\bm k'_e$ and $\bm k'_o$ are given by $¸{\bm k'}_\parallel$ of the incident plane wave, as discussed before. Their $z$-components 
are determined by the  refraction indices $n_{ord}, n_{ext}$ given as follows:
\beal
\label{eq:extord}
\bm M[\bm k_o']\cdot \bm o_0[\bm k_o'] &= 0 \;\Rightarrow (\bm k_o')^2 = n_{ord}^2 = n_o^2 \;,
\nonumber\\
\bm M [\bm k_e']\cdot \bm e_0(\bm k_e') &=0 \;\Rightarrow
(\bm k_e')^2 = n_{ext}^2 = n_e^2 - \beta (\bm k_e' \cdot \nv)^2 = 0 \;, \; \textrm{with} \; \beta = \frac{n_e^2}{n_o^2} -1.
\end{align}
The corresponding magnetic field vectors  are easy to calculate, since they are perpendicular to $\bm e_0[\bm k'_e], \bm k'_e$ and  to $\bm o_0[\bm k'_o],  \bm k_o'$, respectively (see Eq.~(\ref{eq:maxall})).

To determine the amplitudes of the diffracted and reflected plane waves in dependence 
on the wavevector and the polarization of the incident plane wave 
one makes use of the
continuity of the tangential components of the electric and magnetic fields at any interface. In addition
the normal components of the displacement $\bm D$ and the magnetic induction $\bm B$ have to be continuous as well.  The calculational details are presented in Appendix~\ref{sec:reflect}.

\section{The short-wavelength approximation for the optics of materials with uniaxial symmetry}
\label{sec:shortwave}
As demonstrated in Sec.~\ref{subsec:light} the general monochromatic Maxwell equations (\ref{eq:el}), (\ref{eq:mag})
can easily be solved in terms of plane waves for a constant dielectric tensor $\bm \varepsilon$.   We are 
now interested in the case where $\bm \varepsilon$ varies 
slowly in space on a scale $2\pi/q$ which is much larger than the wavelength $\lambda = 2 \pi/{k_0}, k_0 = \omega/c$ of the light wave in vacuum. 
The position dependence of  $\bm \varepsilon$ is induced by the spatial variations of $\nv$.

As described for instance in \cite{Born} (chapter III), it is appropriate to   solve the monochromatic Maxwell equations (Eqs.~(\ref{eq:maxs})) by using the ansatz:
\be
\label{eq:eikborn}
\bm E(\bm r) = \bm {E}_0 (\bm r) e^{\im k_0 S(\bm r)} \;, \;\;
\bm H(\bm r) = \bm {H}_0 (\bm r)  e^{\im k_0 S(\bm r)}. 
\ee
The {\it eikonal} $S$ is a real scalar function of position, while the field amplitudes $\bm  E_0, \bm {H}_0$ are in general complex vector functions of position.
The ansatz for $\bm E(\bm r)$ in Eq.~(\ref{eq:eikborn}) is now inserted into Eq.~(\ref{eq:el})
where we make use of the general identity:
\be
\label{eq:iden}
\bm \nabla \times  (\bm \nabla \times  (\bm V \exp[i k_0 S)]) = \exp[i k_0 S]\,
(\bm \nabla + i k_0 \bm \nabla S) \times  [ (\bm \nabla +ik_0  \bm \nabla S) \times  \bm V)]
\ee
for an arbitrary vector $\bm V$ and a scalar S, which are both  space dependent.
Collecting the terms $O(k_0^2)$ we arrive at the eikonal equation:
\be \label{eq:S}
\left [(\bm \nabla S)^2 - \bm \nabla S  \otimes \bm \nabla S  - \mu (\epsilon_\perp + (\, \epsilon_\parallel -\epsilon_\perp) \nv \otimes \nv \, ) \, \right] \cdot \bm E_0 = 0.
\ee
Comparison with Eq.~(\ref{eq:ME}) shows that $\bm \nabla S$ corresponds  to a local version of the (dimensionless) wavevector
$\bm k'$ in the case of  constant $\nv$. Accordingly we introduce local representations $\bm o^N$ and 
$\bm e^N$   of the ordinary 
and extraordinary polarization  vectors $\bm o_0$, $\bm e_0$ defined in Eq.~(\ref{eq:veco}):
\be \label{eq:vecoN}
\bm o^N(\bm r)  = \frac{\bm \nabla S_o(\bm r) \times  \nv(\bm r)}{|\bm \nabla S_o(\bm r) \times  \nv(\bm r)|},\quad
\bm e^N(\bm r) = \frac{\epsilon_\perp \nv(\bm r) - \bm \nabla S_e(\bm r) (\bm \nabla S_e(\bm r) \cdot \nv(\bm r) )}{|\epsilon_\perp \nv(\bm r) - \bm \nabla S_e(\bm r) (\bm \nabla S_e(\bm r) \cdot \nv(\bm r) )|},
\ee
the position dependence of which is often  suppressed in the following.    
Thus we have to distinguish between  the ordinary eikonal solution, $S_o$, and
the extraordinary one, $S_e$,  of Eq.~(\ref{eq:S}), which correspond 
to the solutions $\bm E_0 = \bm o^N$ and  $\bm E_0 = \bm e^N$ in 
Eq.~(\ref{eq:vecoN}). 
Obviously  the expressions   for $\bm k'_e$,  $\bm k'_o$ in Eq.~(\ref{eq:extord}) translate immediately into the following
equations for the local wavevectors $\bm \nabla S_o$ and  $\bm \nabla S_e$:
\bseq
\label{eq:eikgen}
\beal
\label{eq:eikso}
(\bm \nabla S_o)^2 &= n_o^2,\\
 \label{eq:eikse}
(\bm \nabla S_e)^2 +  \beta (\bm  S_e \cdot \nv )^2 &= n_e^2.
\end{align}
\eseq
The solutions $S^0_e, S^0_o$ of Eqs.~(\ref{eq:eikgen})
for the homogeneous basic planar state where $\nv = \nv_0 = \bm{\hat x}$, which 
are associated with  a plane wave with wavevector $\bm k' = (k'_x, k'_y, k'_z) \equiv \bm k'_\parallel + k'_z \bm e_z$, 
 read as follows: 
\be 
\label{eq:Sbas}
S^0_o(\bm x, z) = \bm k'_\parallel \cdot  \bm x +  {k'_z}^o( \bm k')\, z \;, \;\;
S^0_e(\bm x, z) = \bm k'_\parallel \cdot  \bm x +  {k'_z}^e( \bm k')\, z \;,
\ee
with
\be
\label{eq:kz}
 {k_z'}^o(\bm k') = n_o\Big[1 - \frac{{k'_x}^2 + {k'_y}^2}{n_o^2}\Big]^{1/2} \;, \;\;
 {k_z'}^e(\bm k') = n_e \Big[1 - \frac{{k'_x}^2}{n_o^2}
- \frac{{k'_y}^2}{n_e^2}\Big]^{1/2}.
\ee

It is convenient to split off from  the eikonal $S_e$ in Eq.~(\ref{eq:eikse}) the contribution  $S^0_e$ in Eq.~(\ref{eq:Sbas}) of the basic state by using the ansatz  $S_e = S^0_e + \bar S_e$.   
Thus one arrives at a modified eikonal equation for $\bar S_e$
as a  quadratic form in the partial derivatives $\partial_z \bar S_e$: 
\be \label{eq:eikz}
A(\frac{\partial}{\partial z} \bar S_e)^2 + 2 B (\frac{\partial}{\partial z} \bar S_e) -C = 0
\;, \;\; \bar S_e = S_e - S_e^0.
\ee
The coefficients $A, B$ and $C$ depend on the derivatives   $\partial_x \bar S_e, \partial_y \bar S_e$ and on the director  distortion $\Delta \bm n$.
In analogy to \cite{Pana}  Eq.~(\ref{eq:eikz}) can be rewritten as:
\be \label{eq:eikzwur}
\frac{\partial}{\partial z} \bar S_e = \frac{-B + \sqrt{B^2 +AC}}{A} =  \frac{C}{B + \sqrt{B^2 +AC}}.
\ee  
Equation (\ref{eq:eikzwur}) has to be solved with the initial condition 
$ \bar S_e(\bm x, z =0) =0$ since $S_e(\bm x, 0) \equiv 
 S^0_e(\bm x,  0)$, which  is already split off in Eq.~(\ref{eq:eikz}). Note that for the same reason we had to exclude the negative  square root 
in the transition from Eq.~(\ref{eq:eikz}) to Eq.~(\ref{eq:eikzwur}). 
 
For the director distortion $\Delta \bm n$ given in Eq.~(\ref{eq:Delncomp}) 
we restrict ourselves  to  solutions of Eq.~(\ref{eq:eikzwur}) up to second order in the small quantities $\theta_m$ and $|\bm k'_\parallel|$, i.e., allowing only for small deviations 
from  perpendicular incidence ($|\bm k'_\parallel| =0$).     
Thus we  represent  $\bar S_e$ in the form:
\be
\label{eq:Sexpan}
\bar S_e(\bm x,z) =  \theta_m  S^{(1,0,0)}(\bm x,z) + k'_x \theta_m S^{(1,1,0)}(\bm x,z) +  k'_y \theta_n S^{(1,0,1)}(\bm x,z)
 + \theta_m^2\bar S^{(2)}(\bm x,z) + \cdots
\ee
Note that $\theta_m$-independent terms do not appear; they are already covered by 
 $S_e^0(\bm x, z)$ (see Eq.~(\ref{eq:Sbas})).  
To proceed it is sufficient to keep in the coefficients $A, B, C$ of Eq.~(\ref{eq:eikzwur}) only  the terms contributing to the expansions coefficients of $\bar S_e$ in   Eq.~(\ref{eq:Sexpan}).
After simple algebra we thus arrive at the following leading terms which depend on the 
components of the director distortion $\Delta \bm n$ (see Eq.~(\ref{eq:Delncomp})) and on the 
components of $\bm k'$ as follows:
\be
\label{eq:eikfinAB}
A =1 \;,
\quad B = {k'_z}^e  + \beta \theta_m \delns_z k'_x \;, \quad
C = -\beta n_e^2 \theta^2_m (\delns_z)^2 - 2 \beta k'_x n_e  \delns_z.
\ee
Using these expressions we obtain the following ordinary differential equations   for $S^{(1,1,0)}(\bm x,z)$ and $S^{(2)}(\bm x,z)$:
\be
\label{eq:S2ord}
\frac{\partial S^{(1,1,0)}(\bm x,z)}{\partial z} = -  \beta \delns_z(\bm x,z) \;, \quad
\frac{\partial  S^{(2)}(\bm x,z)}{\partial z} = -\frac{\beta}{2} n_e \, (\delns_z(\bm x,z))^2 \;, 
\ee
while $ S^{(1,0,0)} \equiv 0$ and $S^{(1,0,1)} \equiv 0$. Note, that  
 neither a twist of the director in the plane ($\delns_y \ne 0$) nor finite values of $k'_y$ are reflected in $\bar S_e$ in this order. 
According to Eq.~(\ref{eq:S2ord}) the calculation of the eikonal solution  requires only  
$z$-integrations.  
Choosing for $\delns_z (\bm x,z)$ in particular the one-mode representation in 
 Eq.~(\ref{eq:ansatz_n}) we obtain in this way: 
\be
\label{eq:Setot}
\bar S_e(\bm x, z) = - \theta_m d \frac{\beta}{4 \pi} \cos(\bm q \cdot \bm x)\Big( 8 k'_x \sin^2(\pi z/{2 d})  + \theta_m \frac{n_e}{2} (\frac{2 z \pi}{d}  - 
\sin(2 \pi z/d))  \cos (\bm q \cdot \bm x)\Big) \;. 
\ee
In Appendix~\ref{sec:eikonaln} we discuss  the general solution of Eq.~(\ref{eq:eikzwur})  up to orders $O(\theta_m^2)$ and $|\bm k'_\parallel|^2$, where the impact of finite $k'_y$
and $\delns_y$ becomes visible.

In a next step we have to determine the amplitude $\bm E_0$  (Eq.~(\ref{eq:eikborn}))  of the 
electric field, which is expanded in terms of  $\bm o^N$ and $\bm e^N$ and a third 
linearly independent vector, $\bm s^N$. Following  \cite{Pana}
$\bm s^N$ is defined as follows:
\be
\label{eq:sNdef}
\bm s^N = \frac{\bm \nabla (S_o + S_e)}{|\bm \nabla (S_o + S_e)|}, \; \text{thus} \;  \bm s^N = \bm{\hat z} \; \text{for} \; \bm k'_\parallel =0, 
\ee
and the electric field is finally represented as:
\be 
\label{eq:eik0}
\bm E(\bm r) = \mathbb{O}(\bm r)   \bm o^N(\bm r) e^{\im k_0 S_o} + \mathbb{E}(\bm r) \bm e^N(\bm r) e^{\im k_0 S_e} + \im \mathbb{Z}(\bm r) \frac{\bm s^N(\bm r)}{k_0}
e^{\im k_0 (S_e + S_o)/2} \;,
\ee
with amplitudes $\mathbb{O},  \mathbb{E}$ and   $\mathbb{Z}$ and the eikonal solutions  $S_o(\bm r), S_e(\bm r)$ of Eqs.~(\ref{eq:eikgen}).  
Introducing the  ansatz Eq.~(\ref{eq:eik0}) into Eq.~(\ref{eq:el}) and 
making use  of Eq.~(\ref{eq:iden}) leads eventually to coupled partial differential equations for 
$\mathbb{O}(\bm x,z), \mathbb{E}(\bm x,z)$ presented in Appendix~\ref{sec:ampliE}. 
The equations are solved iteratively
as power series in  $\theta_m$, where 
the power series expansion of the extraordinary eikonal $\bar S_e(\bm x,z)$ (see Eq.~(\ref{eq:Setot})) serves as input. 
Some explicit results will be given in Secs. \ref{sec:short2d}, \ref{sec:3dpat} and  discussed in Sect. \ref{sec:discuss}. In the following section, however, it will be demonstrated that the determination of the amplitudes  $\mathbb{O}, \mathbb{E}$ can be circumvented in the frequently used two-dimensional experimental geometry.
  
It should be emphasized, that in the approach used in \cite{cannell} the steps of our procedure are just reversed. First the field amplitudes 
are determined by using Fermat's principle to determine the light ray trajectories, which is technically cumbersome. Then the eikonal $S_e$ is obtained by 
summing up the phases along the ray path.

\section{The short-wavelength approximation in a 2D geometry}
\label{sec:short2d}

In this section we restrict ourselves to the special case of an incident TM wave, where the wavevector $\bm k'$ and the electric field amplitude $\bm E_0$ of  the plane wave propagating  into the nematic layer
are confined to the $xz$-plane.  Furthermore the director distortion
$\deln$ defined in Eq.~(\ref{eq:ansatz_n}) is assumed to be parallel to this plane as well. Finally we consider only stripe pattern (called ``normal'' rolls  in EC) where   the wavevector $\bm q$  characterizing $\deln$ in Eq.~(\ref{eq:Delncomp})  is parallel to $\bm n_0 = \bm {\hat x}$, the director orientation in  the basic state. Thus all quantities 
depend only on $x,z$. 
Inspection of the Maxwell equations reveals, that 
$\mathbb{E}$ remains in the $xz$-plane inside the nematic layer, i.e. $\mathbb{O}(x,z) \equiv 0$.   Only this  exclusively extraordinary configuration has been discussed in the literature \cite{cannell,stann,rasenat89} so far.

Instead of analyzing equations for the electric field (see Eqs.~(\ref{eq:ampE}) in Appendix~\ref{sec:ampliE}) we approach the problem from a slightly different perspective. 
In the 2D-geometry the magnetic induction $\bm B$  has only a  non-vanishing $y-$component, $B_y$,   and it is expected that an analysis based on the y-component of Eq.~(\ref{eq:mag}), a scalar equation, is more transparent and straightforward  than the analysis based on the electric field. First of all  we use the fact 
that in the present geometry Eq.~(\ref{eq:mag}) leads to  the following
scalar equation for $B _y(x,z)$:
\be \label{eq:bzeng}
\bm \nabla \cdot (\bm \varepsilon \cdot  \bm \nabla B_y) + \epsilon_\parallel
\epsilon_\perp \mu k_0^2 B_y = 0.
\ee
Instead of solving Eq.~(\ref{eq:bzeng}) by using a representation of  $B_y$  with a complex 
phase as in \cite{Zengin1}   
 we follow closely the approach discussed in Sec.~\ref{sec:shortwave} using  the ansatz: 
\be
\label{eq:Beikan}
B_y(x,z) =  \bar B(x,z) \exp [i k_0 S_e(x,z)], 
\ee 
which requires much less effort.
We arrive again at the eikonal equation (\ref{eq:eikse}) for $S_e$ discussed before in Sec.~\ref{sec:shortwave} and to the following equation for $\bar B(x,z)$:  
\be
\bm \nabla  S \cdot (\bm \varepsilon \cdot \bm \nabla \bar B(x,z)) + \bm \nabla [\bar B(x,z)
(\bm \varepsilon \cdot \bm \nabla \bar S_e)] = 0 \label{eq:eikob},
\ee
with $\bar S_e = S_e -S_e^0$ (see Eq.~(\ref{eq:eikz})).

We will concentrate on the case of small  $\theta_m$ and small $k'_x$ and use  the following expansion   for the  magnetic field amplitude $\bar B(x,z)$ (\ref{eq:eikob}): 
\be \label{eq:Bexpand}
\bar B(x,z) = B_0  +  \theta_m   B^{(1)}(x, z) +
\theta^2_m B^{(2)}(x, z) + \theta_m \, k'_x  B^{(1,1)}(x, z) .
\ee
To order $\theta_m$  Eq.~(\ref{eq:eikob}) reduces to: 
\be \label {eq:blin}
2 \frac{\partial}{\partial z} B^{(1)}(x,z) = - \beta B_0  \frac{\partial}{\partial x} \delns_z(x,z),
\ee
which has to be solved with the initial condition  $B^{(1)}(x,z=0) = 0$.
When using again the one-mode approximation for $\delns_z(x,z)$ in  Eq.~(\ref{eq:ansatz_n}) we arrive
by direct integration at
\be \LB{eq:bsolveone}
B^{(1)}(x,z) =  B_0 q d \frac{\beta}{2 \pi}  \sin (q x ) (1 - \cos(\pi z/d)).
\ee
The higher order amplitudes defined in Eq.~(\ref{eq:Bexpand}), which give also contributions
to the amplitude grating efficiency of the distorted nematic layer, are given in Appendix~\ref{sec:mag2}. For small  director distortion amplitudes $\theta_m$, considered in this work, they are much smaller than the corresponding phase grating terms $\propto k_0 \bar S_e(x,z)$ (see Eq.~(\ref{eq:Setot})). Thus in the  following only the leading amplitude term $\propto \theta_m$ like $B^{(1)}$ will be  kept. 
 
The expansion coefficients  of the total magnetic field amplitude  $\bar B(x,d)/B_0$ defined in Eq.~(\ref{eq:Bexpand})  agree with those given in \cite{Zengin1}.  Furthermore they allow 
for the calculation of the electric field by using the general 
Maxwell equations (\ref{eq:maxall}). In fact we arrive in this way  
at the electric field amplitudes $\mathbb{E}(x,d)/\mathbb{E}_0$ (see Eq.~(\ref{eq:eik0})), 
as calculated in \cite{cannell} by considerably more intricate 
calculations. For completeness we have convinced ourselves that also the direct solution of Eqs.~(\ref{eq:ampE}) leads  to the same result. This perfect agreement (for more details see \cite{can_compare})  
serves as a most convincing test of our procedure.

\section{ Optical analysis of general 3D pattern}
\label{sec:3dpat}
In the following we describe the optical properties  of general 3D configurations in nematics. First the wavevector $\bm q$  characterizing  the periodic distortion
$\deln$ (see Eq.~(\ref{eq:ansatz_n})) may not be parallel to $\bm n_0 = \bm{\hat x}$. 
Furthermore both the  electric field polarization  
and the orientation of the wavevector $\bm k = k_0 \bm  k'$ of the incident plane wave  
are in principle arbitrary. 
In general we use  the following representation for the 
(dimensionless) wavevector  $\bm k'$ of the plane wave, which enters the nematic layer from a glass plate (refraction index $n_g$):
\be 
\label{eq:wavedef}
\bm k' =  \bm k'_\parallel + k'_z \bm {\hat z} = \Big(\sin (\vartheta_g) \cos \phi,\, \sin (\vartheta_g) \sin \phi, \, \sqrt{n_g^2- \sin^2 (\vartheta_g)}\Big)\,.
\ee
Thus $|\bm k'|^2 = n_g^2$ (see Sec.~\ref{subsec:light})  is automatically guaranteed. The polar angle $\vartheta_g$ describes the inclination of the incoming ray with respect to the layer normal (parallel to $\bm {\hat z}$) and the azimuthal angle, $\phi$, a rotation of $\bm k'$ about this axis.  It is obvious that  the angles $\vartheta_g$ and $\phi$ can be alternatively interpreted in terms of a tilt  and  a rotation of the nematic layer  at fixed $\bm k'$. 
In line with the typical experimental setup 
we will restrict ourselves in the following to  small  $|\bm k'_\parallel|$, i.e. to small $\vartheta_g$. Then up to order $O(\sin(\vartheta_g))$ 
the wavevectors of the extraordinary and the ordinary waves inside the nematic layer 
are given as $
\bm k' = (\bm k'_\parallel, n_e)$ and $\bm k' = (\bm k'_\parallel, n_o)$, respectively (see Sec.~\ref{subsec:light}). 

Let us start with a ``toy model'' of a periodic twist modulation of the director  confined to the $xy$-plane of the following form:
\be
\label{eq:twist}
\nv = (\cos \alpha(x,z), \sin \alpha(x,z),0)
\ee
with the twist angle $\alpha(x,z) = \alpha_{m} \vartheta(z) \cos( q x)$ where $\vartheta(z)  = 0$ for $z =0, d$. In the absence of $x$-variations  this director configuration
is realized in the planar geometry as the result of a twist Freedericksz transition, when  a static magnetic field
with amplitude $H$ above a critical threshold field $H_c$ is applied along  the $y$-direction.  The maximal twist amplitude $\alpha_{m}$ varies then like $(H^2 - H^2_c)^{1/2}$. An additional $x$ variation is for instance characteristic for the so called chevron patterns
in EC \cite{ross_physd}, when the applied ac-voltage is turned off after some time.

We consider the special case of an incident TM wave in the $xz$-plane,  which leads to an extraordinary wave with amplitude 
$\mathbb{E}_0$ at the lower surface $z =0$ of the nematic layer (see Eq.~(\ref{eq:eik0})).  
At  $z=d$ we will find a  TM wave as well, since the polarization of a  plane wave follows  adiabatically (Mauguin principle) the orientation of $\bm e^N(x,z)$ (\ref{eq:vecoN}) and is thus again parallel to $\bm {\hat x}$ at $z =d$. Due to the twist in the director field, however,  a TE electric field 
component is observable as well in the upper glass plate, since also an ordinary wave with amplitude $\mathbb{O}(x,z)$ develops in the layer.  To leading order in $\alpha_{m}, k_x$ we obtain 
easily from Eqs.~(\ref{eq:ampE}) the following partial differential equation 
for $\mathbb{O}(x,z)$:
\be
\label{eq:Eotwi}
\big(n_o \frac{\partial}{\partial z}  + k_x \frac{\partial}{\partial x}\big) \mathbb{O}(x,z)    = e^{\im k_0 (n_e -n_o) z}
 \mathbb{E}_0 \big (n_e  \frac{\partial}{\partial z} + k_x \frac{\partial}{\partial x}\big) 
\cos (q x) \vartheta (z). 
\ee
The solution of Eq.~(\ref{eq:Eotwi}) in Fourier space by using the ansatz
\be 
\mathbb{O}(x,z) = \mathbb{O}'(q,z) \exp[\im q x] 
+  \mathbb{O'}(-q,z) \exp[-\im q x],
\ee
reads as follows:   
\be
\mathbb{O}'(x,\pm q) = \mathbb{E}_0  \alpha_{max} \frac{n_e}{2 n_o}  
\exp[\pm \im q k_x' z]\int_0^z dz' \exp [\im k_0(n_e -n_o  \mp k'_x q) z] \big (\partial_{z'} \pm i k'_x q\big) \vartheta (z').
\ee
For perpendicular incidence ($k'_x = 0$) we obtain thus: 
\be
\label{eq:Eo1twi}
\mathbb{O}(x,z) = \mathbb{E}_0 \alpha_{m} \frac{n_e}{n_o} \cos(q x) \int_0^z dz' e^{\im k_0(n_e -n_o) z'} \frac{d}{d z'} \vartheta (z').
\ee
This expression  has been first derived  in \cite{schadt80} for $q =0$ by a different method. There it has been emphasized that the electric field
intensity $|\mathbb{O}(z =d)|^2$,  which is singled out and measured by using 
crossed polarizers at $z =0,d$,  gives   valuable informations on the elastic constants of nematics. Note also  the sensitive dependence of  $|\mathbb{O}(z =d)|^2$  on $\omega, d$ via the large exponent $k_0 d(n_e -n_o)$. 
It should be mentioned  that the amplitude $\mathbb{O}(x,z)$ in Eq.~(\ref{eq:Eo1twi}) 
has been also derived in \cite{amm98} by using the Jones matrix method, except that 
the prefactor $n_e/n_o$ is approximated  by one.

After this introductory  considerations we will now deal with the general and mostly  realized case of director distortions which contain also a tilt contribution (see Eq.~(\ref{eq:ansatz_n})). The corresponding eikonals, $S_o, S_e$,  
have been already given in Sec.~\ref{sec:shortwave} (see Eqs.~(\ref{eq:Sbas}, \ref{eq:Setot})). 
To determine the field amplitudes $\mathbb{E}(\bm x, z),\mathbb{O}(\bm x, z)$ defined in Eq.~(\ref{eq:eik0})
we have to solve Eqs.~(\ref{eq:ampE}).
We restrict ourselves to solutions up to order $\theta_m$ and may neglect as discussed in Appendix~\ref{sec:mag2} the dependence of the amplitudes on $|\bm k'_\parallel|$. Thus the field amplitudes are expanded as follows:
\be \label{eq:fieldexpan}
\mathbb{O} = \mathbb{O}_0 + \theta_m \mathbb{O}^{(1)}(\bm x,z), \;
\mathbb{E} = \mathbb{E}_0 + \theta_m \mathbb{E}^{(1)}(\bm x,z).
\ee
To first order in $\theta_m$ we arrive from Eqs.~(\ref{eq:ampE}) easily at the following partial differential equations for
$\mathbb{O}^{(1)},  \mathbb{E}^{(1)}$ in terms of the components of $\Delta \bm n $ (see Eq.~(\ref{eq:Delncomp})):
\bseq \label{eq:Eode}
\begin{align}
\frac{\partial}{\partial z} \mathbb{E}^{(1)}(\bm x,z) &= - \mathbb{E}_0 \frac{\beta}{2}  
\frac{\partial}{\partial x} \delns_z(\bm x,z)-  \mathbb{O}_0 \frac{a_{-}^2 n_o}{n_e} 
\frac{\partial}{\partial z} \delns_y(\bm x ,z) ,\label{eq:Eode1}\\
\frac{\partial}{\partial z} \mathbb{O}^{(1)}(\bm x , z)  &= \mathbb{E}_0 \frac{ a_{+}^2 n_e}{ n_o}
\Big[\frac{\partial}{\partial z} \delns_y(\bm x ,z) + \frac{\beta}{2} \frac{\partial}{\partial y} \delns_z(\bm x ,z)\Big].
 \label{eq:Eode2}
\end{align}
\eseq
According to Eqs.~(\ref{eq:ampE}) the functions $a^2_{\pm}(\bm x ,z)$ are given as:  
\be \label{eq:aplumindef}
a^2_{\pm}(\bm x ,z) = \exp \left [\pm \im  k_0 \left((n_e - n_o ) z + \bar S_e(\bm x,z)\right) \right],
\ee
where $\bar S_e$ within the one-mode approximation can be found  in Eq.~(\ref{eq:Setot}).
Inspection of the transmission coefficients in Eq.~(\ref{eq:transcoef}) shows  
that  the initial ordinary and extraordinary electric field amplitudes $\mathbb{O}_0,  \mathbb{E}_0$ at $z =0$ can be realized by using the incident electric field
as a  superposition of a TM and a TE wave.

Equations (\ref{eq:Eode}) are easy to solve and we obtain:
\bseq \label{eq:efieldfin}
\bal
\mathbb{E}^{(1)}(\bm x,z) =  \mathbb{E}_0 \Big [1  -  \frac{\beta}{2} \int_0^z dz'  \frac{\partial}{\partial x} \delns_z(\bm x ,z')\Big] -  \mathbb{O}_0 
\frac{n_o}{n_e} \int_0^z dz'  a^2_{-}(\bm x ,z')
\frac{\partial}{\partial z} \delns_{y}(\bm x ,z'),\label{eq:efieldfine}\\
\mathbb{O}^{(1)}(\bm x,z) =   \mathbb{O}_0 + \mathbb{E}_0\frac {n_e}{n_o} 
\int_0^z dz'  a^2_{+}(\bm x,z')
\left[ \frac{\partial}{\partial z} \delns_y(\bm x,z') + \frac{\beta}{2} \frac{\partial}{\partial y} \delns_z (\bm x ,z')\right] \label{eq:efieldfino}.  
\end{align}
\eseq
The final expressions in Eqs.~(\ref{eq:efieldfin}) are not difficult to interpret.
First of all the extraordinary field amplitude $\mathbb{E}^{(1)}$,  obtained before in the 2D geometry (see Eq.~(\ref{eq:blin})) where $\mathbb{O}_0 = 0$ and $\bm q =(q, 0)$,  is recovered, as it should be. 
Of particular importance is the fact that according to Eq.~(\ref{eq:efieldfine}) a finite incident ordinary field amplitude  $\mathbb{O}_0$ at $z =0$  is sufficient to generate an extraordinary field component for finite $z$ and thus also at $z =d$. Analogously we obtain 
in Eq.~(\ref{eq:efieldfino}) from a nonzero $\mathbb{E}_0$ a finite ordinary field amplitude  $\mathbb{O}(\bm x,d)$.   
In close analogy  to Eq.~(\ref{eq:Eo1twi}) one needs in any case  a twist of the director field
($\delns_y \ne 0$) or  a finite angle between the wavevector $\bm q$ and $\bm n_0$ (leading to a $y$-dependence of $\delns_z$, see Eq.~(\ref{eq:ansatz_n})).  
Note, that given the amplitudes $\mathbb{E}^{(1)}$ and  $\mathbb{O}^{(1)}$
the  full  extraordinary and ordinary electric fields are obtained according
to Eq.~(\ref{eq:eik0}) by a multiplication with the corresponding phase factors and the 
polarization vectors. The detailed discussion is postponed to 
Sec.~\ref{sec:discuss}. 
  
We have demonstrated in this section and before in Sec.~\ref{sec:short2d}, that 
the electromagnetic field at the upper surface $z =d$ of the nematic layer
shows periodic variations with wavevector $\bm q$ both in the amplitude
(amplitude grating) and in the the phase (phase grating). 
Using these data the field 
in the isotropic media (glass, air) above the nematic layer has to be constructed by solving the isotropic Maxwell equations with constant dielectric constants. 
This task is formally easy  by expanding  the fields at $z =d$ into Fourier series in terms of $e^{\im n \bm q\cdot \bm x}, n =\pm 1, \pm 2 \dots$, i.e. the field is represented as a superposition of plane waves.  The whole procedure and its implications have been discussed in \cite{cannell} for   RBC. 
The corresponding analysis for nematics is presented in Appendix~\ref{sec:talbot}.
The moduli of the Fourier components give directly the intensity of the diffraction fringes. The interference of the plane waves produces the 
shadowgraph picture.  The main results are collected and discussed
in the following section.

\section{Summary and Discussion}
\label{sec:discuss}
In the previous sections we have analyzed the optical properties of a thin nematic layer
of thickness $d$ in the presence of small director  perturbations  $\Delta \bm n (\bm x, z)$ 
(Eq.~(\ref{eq:Delncomp})) of the  planar basic state ($\bm n_0 = \bm{\hat x} , 0 \le z \le d$).  The perturbation is periodic
in the $xy$-plane with wavevector $\bm q$. The optical properties of the layer are probed using a monochromatic plane wave with frequency 
$\omega$ and wavevector $\bm k$ (see Eq.~(\ref{eq:wavedef})) which enters the layer at $z  = 0$ from a glass plate with refraction index $n_g$.
The goal of this section is to summarize and to discuss the main results on the optical analysis, which have been described in Secs. \ref{sec:short2d}, \ref{sec:3dpat} as well as in several Appendices. 

 The incident wave can in general be decomposed
into a superposition of a TE and TM wave with polarization vectors $\bm p_E, \bm p_M$ defined in Appendix~\ref{sec:reflect}.
Using the transfer coefficients given there  the amplitudes 
$\mathbb{O}_0$ and  $\mathbb{E}_0$ of the ordinary and extraordinary waves just at the lower surface  
at $z = 0$ of the nematic layer are directly given. Inside the nematic layer ($0 \le z \le d$) the 
electric field is represented by  an amplitude and a phase factor as discussed in  Sec.~\ref{sec:shortwave}. Let us first concentrate on the extraordinary waves which are almost exclusively utilized  in EC experiments, where $\bm q$ includes at most a moderate angle with 
$\bm n_0 \parallel \bm {\hat x}$. Furthermore only the leading terms in  inclination angle $\vartheta_g$ 
(Eq.~(\ref{eq:wavedef}))  of the incident plane wave are  kept.  
In this approximation the extraordinary electric field $\bm E_e(\bm x)$ at $z =d$ given in Eq.~(\ref{eq:fieldn1}) reads as:  
\be
\label{eq:fieldn}
\bm E_e(\bm x) = \mathbb{E}_0 \exp{[\im k_0 (\bm k'_\parallel \cdot \bm x + n_e d)]} \sum_{n = -\infty}^{\infty} C^N(n) e^{i\,n \scal{q}{x}}\,
\bm e_0[\bm k'_\parallel].
\ee
The Fourier coefficients $C^N(n)$  describe the optical grating characteristics of the nematic fluid layer at $z =d$ and eventually also the intensity of the TM wave 
in air above the nematic layer (see Eq.~(\ref{eq:represen})).   

The  intensity of the two diffraction spots of order $n$ 
relative to the intensity $|\mathbb{E}_0 T_{eg} T_{ga}|^2$ of the transmitted wave (see Eq.~(\ref{eq:represen})) is  in general given as $|C^N(\bm n)|^2$. The diffraction spots  characterized by the wavevectors $\bm k = (k_0 k' \pm n \bm q) \bm{\hat  x} + k_0 \bm {\hat z}$, are observed in directions which enclose  small angles $\gamma_{\pm n}$ with $\bm{\hat z}$ given by:
 \be
\label{eq:gamn}
\gamma_{\pm n} = \arctan \left [\frac{|k_0 \bm k'_\parallel \pm \bm n q|}{k_0}\right ],
\ee   
where we have neglected the terms of the order $O(n^2 q^2/{k_0}^2)$.

Let us start with the contribution of the leading Fourier coefficients $C^N(\pm 1)$ .
The relative intensities of the diffraction spots are determined by $|C^N(\pm 1)|^2$, which 
 according to Eq.~(\ref{eq:Ccoef}) are given as follows:
\be 
\label{eq:fring1} 
|C^N(\pm 1)|^2 = \frac{\theta_m^2}{4}  |c_{E1}|^2 |1 \mp \frac{c_{S1}}{c_{E1}}|^2, \quad \text{with} \quad  \frac{c_{S1}}{c_{E1}} = -\frac{2 k_0}{q_x} \sin(\vartheta_g) \cos \phi.    
\ee 
Here we have inserted  the expressions for the coefficients $c_{S1}, c_{E1}$ in Eq.~(\ref{eq:expcoef}). 

At first  a small asymmetry between $n =1$ and $n =-1$  in the case of oblique incidence ($\vartheta_g \ne 0 $) is evident. The asymmetry is reflected  both in the angles $\gamma_{\pm 1}$ (Eq.~(\ref{eq:gamn})) and in the intensities (Eq.~(\ref{eq:fring1})).     
A closer look at the intensities  reveals a competition between the amplitude grating ($c_{E1})$ and phase grating coefficients ($c_{S1}$). For $\phi =0$ 
already at the  small inclination angle  $\vartheta_g = q_x/(2 k_0)$ phase grating starts to prevail.    
For standard  EC experiments  and medium ac frequencies  $\Omega$ where  typically $q d \approx 1.5 \pi$    we obtain  thus $\vartheta_g  \approx 0.012  \approx 0.7^\circ$  when using visible light with $\lambda = 0.6 \mu m$  and a layer  thickness of $20 \mu \mathrm{m}$. 
Consequently for an angle of  $7^\circ$, say,  the intensity 
$|C^N(1)|^2$  is larger by a factor $100$ compared to perpendicular incidence ($\vartheta_g =0$).   
For smaller $\Omega$  in EC often oblique rolls appear; the angle between $\bm q$ and $\bm n_0$
becomes finite, while the modulus of $\bm q$ remains practically unchanged. Since thus
the $x$-component, $q_x$, of the wavevector $\bm q$ decreases the denominator of the ratio $c_{S1}/c_{E1}$ (Eq.~(\ref{eq:fring1})) gets larger and phase grating becomes
dominant at even smaller $\vartheta_g$.  On the other hand an azimuthal rotation of the  incidence plane of the incident light ($\phi$ finite) leads to a reduction of the phase grating effect due to the factor 
$\cos \phi$ in the ratio $c_{S1}/c_{E1}$. 
In conclusion, we have demonstrated  that even a small inclination of the incoming ray 
leads to a dramatic increase of the intensity of the first order fringes.
So  far we are only aware of one  systematic investigation of this effect \cite{stann}, where
the main experimental findings have been well  described exclusively on the 
basis of phase grating ($c_{E1} =0$). For completeness, we refer to a later work \cite{decay}, where a small inclination of the nematic has been applied as well to enhance the intensity of the first order fringe.

Let us now consider the 
intensity of the second order fringes ($n =\pm 2$), in particular in relation to the  intensity of the first order fringes.  In the case of perpendicular incidence of the light and $\phi =0$ we have thus  to compare  $|C^N(2)|=  \theta_m^2 n_e k_0 d \beta/16$ (see Eq.~(\ref{eq:Ccoef})) with $|C^N(1)|=  \theta_m q_x d \beta/(2 \pi)$. 
Using as in the previous paragraph the ratio $q_x/k_0 = 1/40$ 
we find that for  $\theta_m > \theta_{m2} = 8 q_x/(k_0 n_e \pi) \approx 0.064/{n_e}$   
the intensities of the second order fringes start to out-compete the first order ones. 
For MBBA where $n_e =1.75$ one has $\theta_{m 2} = 2^\circ$.
Using the relation $\theta_m = 80^\circ  \, \sqrt{\epsilon}$ between the maximal tilt angle $\theta_m$ and the control parameter $\epsilon = (U_0^2 -U_c^2)/U_c^2$ in MBBA mentioned at the end of Sec.~\ref{subsec:pattern} the value of $\theta_{m 2} = 2^\circ$ corresponds to $\epsilon = 0.0007$.
It means that already not too far from threshold the the intensity of the second order fringe becomes considerably larger than the intensity of the first order fringe.
This general trend has been clearly demonstrated in experiments \cite{stann}, where 
the first order fringes are not visible at all in the case of perpendicular incidence. 
When  inclining the layer, however,
we have to compare $\theta_m c_{S1}/2$ with $\theta_m^2 c_{S2}/4$, i.e. $(k_0 d)\beta \sin(\vartheta_g)/\pi$ with 
$\theta_m (k_0 d) n_e \beta/16$. Thus for $\vartheta_g > \theta_m n_e \pi/16$ the first order fringe
prevails. 
For MBBA this is for instance the case for  $ \vartheta_g > 2^\circ$ at $\epsilon = 0.005$.

Let us now briefly analyze the shadowgraph intensity. First we concentrating on the contribution, $I^1_s$,  which is characterized by the wavevector $\bm q$.
According to Appendix~\ref{sec:talbot}  we obtain the following  expression  to order $O(\theta_m)$:
\beal
\label{eq:shad1}
I^1_s(\bm x, z) = (E_0 T_{eg} T_{ga})^2 \Big ( 1 + 2 \theta_m [c_{S1} \cos(\bm q \cdot \bm x)
\sin(\frac{q^2}{2 k_0} z') + c_{E1}\sin(\bm q \cdot \bm x) \cos(\frac{q^2}{2 k_0} z')]\Big )
\end{align}
with $z' = z-(d +d_g)$. At first we observe the well known periodicity of the intensity as function $z$ characterized  
by the Talbot wave length $\lambda_T = 4 \pi k_0/q^2$. For the values $q = 1.5 \pi/d$ and $\lambda = 2 \pi /k_0 = 0.6 \mu m$ we obtain thus
$\lambda_T = 1.1$~mm for $d =20 \mu m$. 
The  Talbot periodicity, in clear contrast to the caustics in geometric optics as  
discussed in  \cite{cannell} in detail,  can be easily demonstrated by focusing a microscope on different $z$ levels. There exists always a contribution to $I^1_s$ due to amplitude
grating ($\propto c_{E1}$) but as discussed before the contrast can be considerably 
increased via the phase grating coefficient ($\propto c_{S1}$) in the case of oblique incidence. 
The Fourier modes $\propto \exp[\pm 2 \im \bm q \cdot \bm x]$ of the shadowgraph intensity, 
$I^2_s$,  
 derive from the terms $O(\theta_m^2)$   
in Eq.~(\ref{eq:Ccoef}). As mentioned before the coefficient $c_{S2}$ gives in general  the dominant contribution. Thus we find in second order:
\be 
\label{eq:shad2}
I^2_s (\bm x, z) = (E_0 T_{eg} T_{ga})^2 \theta_m^2 c_{S2} \cos(2 \bm q \cdot \bm x) \cos(\frac{4 q^2}{2 k_0} z') .
\ee
It has been already stressed before that in the case of perpendicular incidence this terms dominates  the Fourier coefficients 
of $I^1_s$ with respect to $\bm x$ even at small $\theta_m$ in line with the typical experimental observations.

So far we have demonstrated, that the use of obliquely incident light
is certainly very important  to increase the contrast, since we obtain phase
grating contributions with wavevector $\bm q$. One might ask whether 
a rotation of the incident plane about the $z$-axis with an angle $\phi$ might have 
some additional advantages. It is for sure that we have only to analyze 
a possible impact on phase grating since amplitude grating effects 
are much smaller. For this purpose we have given the eikonal solutions
also with respect to the $\phi$-dependence in Appendix~\ref{sec:eikonaln} 
(see Eqs.~(\ref{eq:S1}), (\ref{eq:S2})).
The only term of the order $O(\sin(\vartheta_g))$ is the term $f_c(d) \propto\cos(\phi)$ in Eq.~(\ref{eq:fcsexpb}) (corresponding to $c_{S1}$ in Eq.~(\ref{eq:expcoef})) which is maximal for $\phi =0$. First in order 
$O(\sin^2(\vartheta_g))$ terms $\propto \sin(2 \phi)$ appear. There seems 
to be no chance to disentangle these terms from the other ones in experiment. The same problem appears in the contributions with 
wavevector $2 \bm q$ (see Eqs.~(\ref{eq:thetm2})).
Also here the term
$k_0 d \, f_c^{(2)}(d)$ in Eq.~(\ref{eq:cos2}) (corresponding to $c_{S2}$ in Eq.~(\ref{eq:expcoef})) dominates the other ones. Thus, a rotation of the incidence plane does not yield any advantage.

Finally we come to a more detailed discussion of the expressions for electric field amplitudes derived in Eqs.~(\ref{eq:efieldfin}), which describe a possible rotation of the
polarization vectors.
Let us first consider  the first term $\propto \mathbb{E}_0$ on the rhs of Eq.~(\ref{eq:efieldfine}). The $z$-integral gives a nonzero contribution when the $\bm q$-vector 
has a finite $x$-component $q_x$ and one recovers in the one-mode approximation for $\deln$ (Eq.~(\ref{eq:ansatz_n})) the coefficient  $c_{E1}$ in Eq.~(\ref{eq:expcoef}) and discussed after Eq.~(\ref{eq:fring1}).   
The integrals containing the factors $a^2_{\pm}(\bm x, z)$  in Eq.~(\ref{eq:efieldfin})  can be easily evaluated  in the one-mode  approximation
$\vartheta(z) = \sin(\pi z/d)$ by exploiting $k_0 d \gg 1$.   
We  use the identity:
\be 
a^2_{\pm}(\bm x ,z)) =  
\Big [\pm \im  k_0 ((n_e -n_0
- \beta \delns^2_z(\bm x, z)/2  -k'_x \beta  \delns_z)\Big]^{-1} \frac{\partial}{\partial z} a^2_{\pm}(\bm x,z),  
\ee
where we have explicitly introduced the $z$-derivatives of the eikonal solution in
the exponents of $a^2_{\pm }$ (Eq.~(\ref{eq:aplumindef})) by using Eqs.~(\ref{eq:S2ord}). Then we rewrite the integral in Eq.~(\ref{eq:efieldfino}) with the help of an  integration by parts, where only the term $\propto \partial_z \delns_y(\bm x, z)$ give  a nonzero contribution at the boundaries
$z =0,d$. Thus we arrive at: 
\beal
\label{eq:efieldfinex}
\mathbb{E}^{(1)}(\bm x,d) &=  \mathbb{E}_0 \Big [1 +  \frac{\beta}{\pi} (q_x d) \sin (\bm q\cdot \bm x) \Big] -  \mathbb{O}_0 \theta_m a_y  
\frac{n_o}{n_e} \frac{\im \pi (1 + \exp[-\im k_0 \Phi(\bm x)])}{k_0 d (n_e- n_o)} 
\sin(\bm q \cdot \bm x)\nonumber\\
\text{with}\quad  & \Phi(\bm x) = S_e(\bm x, d) = (n_e -n_o)d +  \bar S_e(\bm x, d), 
\end{align}
where $\bar S_e(\bm x,d)$ is explicitly given in Eq.~(\ref{eq:Setot}).
According to Eq.~(\ref{eq:eik0}) the amplitude  $\mathbb{E}^{(1)}(\bm x,d)$ has now to 
be multiplied by the factor  $\exp[i k_0 S_e(\bm x, d)]\, \bm e_0[\bm k']$ to obtain
the extraordinary electric field.
In any case, as long as $\delns_y \ne 0$, we obtain a small amplitude grating term  $\propto \mathbb{O}_0 \theta_m \sin(\bm q\cdot \bm x)(k_0 d)^{-1}$,
where the polarization vector of the incident wave is rotated  by $90^\circ$, when leaving the 
layer. Even for $q_x =0$, but nonzero $\bm k'_\parallel \propto \sin(\vartheta_g)$ and $\mathbb{E}_0$, however, we obtain a much larger phase grating contribution  $\propto (k_0 d) \theta_m \sin(\vartheta_g) \cos(\bm q \cdot x)$, corresponding to the 
coefficient $c_{S1}$ discussed before.   

In a similar manner we may discuss the amplitude  $\mathbb{O}^{(1)}(\bm x, d)$ in Eq.~(\ref{eq:efieldfino}). The integral is evaluated again in the one-mode
approximation as before, where only the term $\propto \partial_z \delns_y$ survives 
in leading order in $(k_0 d)^{-1}$. The total ordinary field is obtained by multiplication of $\mathbb{O}^{(1)}$ with $\bm o_0$ and with the phase factor $\exp[\im k_0 S_o^0(\bm x,d)]$,
where $S_o^0$ (see Eq.~(\ref{eq:Sbas})) does not contain periodic phase modulation terms. Thus only 
the term $\propto \mathbb{E}_0$ leads to optical grating effects and we obtain the total ordinary
electric field amplitude $\bm E_o$ up to order $(k_0 d)^{-1}$ at $z =d$ in the following form:
\beal
\label{eq:ordinary1}
\bm E_o(\bm x, d) = \mathbb{O}^{(1)} \exp[i k_0 S_0]\, \bm o_0[\bm k'] +  \mathbb{E}_0 \theta_m a_y \frac{n_e}{n_o} \frac{\im \pi (1 + \exp[\im k_0 \Phi(\bm x)])}{k_0 d (n_e- n_o)}   \sin(\bm q \cdot \bm x) \bm o_0[\bm k'], 
\end{align}
where  $\Phi(\bm x)$ has been given in Eq.~(\ref{eq:efieldfinex}). In analogy to Eq.~(\ref{eq:efieldfinex}) a twist  in the director field ($a_y \ne 0$) leads again to a rotation
of the incident polarization by $90^\circ$.

The main result of the general discussion of amplitude solutions in Eqs.~(\ref{eq:efieldfin}) is that one should observe even for equilibrium patterns like the flexoelectric domains or the splay-twist Freedericksz pattern ($\bm q \parallel \bm{\hat y}$), diffraction fringes 
and shadowgraph pictures.  This is confirmed in experiments 
\cite{flexo,Meyer, Budanew,flexobook}.
By the way, restricting the light  rays by  apertures or using a slightly divergent light beam, as it happens often in experiments,  amounts in principle also to the generation of obliquely incident rays as well; this effect might be worth to be  analyzed in the future.   
It is needless to say that an analysis based on pure geometric optics 
is unable to capture at all the optical  properties of such patterns.

\section{Concluding remarks }
\label{sec:conclus}

In the previous  sections we have demonstrated that the optics of a nematic layer with a periodically
distorted director field can be convincingly  described by a  standard  short-wave length approximation.  The new calculational scheme is much easier to handle than the previous 
treatments of the problem based on Fermat's principle and the summation of the phases
along the ray path. It is not necessary to distinguish carefully between   phase- and ray refraction indices, which has in fact caused some problems in the earlier work. A big advantage of our new approach is, that  it covers from the beginning three-dimensional experimental geometries  where the polarization of the incident light and the orientation of the wavevector are arbitrary. Thus for instance
the optical analysis of equilibrium patterns with $\bm q \perp \bm n_0$ is put on a firm basis. 
In general it has been demonstrated, that even a slight tilt of the nematic layer leads to a large increase of the intensity contrast of the patterns. 
Note that also a so called pretilt of the director at the confining substrates corresponds to an inclination of the cell.
This feature is based 
on the phase grating effects of patterns in the nematic layer and thus not accessible  in the framework of pure geometric optics.    
We have explicitly discussed only roll solutions characterized by a wavevector $\bm q$.
The generalization to patterns characterized by several wavevectors like squares or hexagons 
is straightforward. Furthermore topological defects in the pattern, like dislocations.
have their counterpart in the corresponding shadowgraph theory as well. It would be certainly a rewarding task  to reconsider  their present, highly sophisticated mathematical description. It is  based on geometric optics and the ensuing  caustics (see e.g. \cite{Joets} and references therein) and should be related to the present approach including 
diffraction effects. 

In this work we have concentrated on the presentation of the theoretical method and in particular on analytical results to leading order in the distortion amplitude of the originally planar
director configuration.  This allows to asses easily the impact of the various experimental 
input parameters.    
A detailed comparison with experimental results, in particular with those presented in \cite{messlin} is planned in near future.

\appendix
\section{Reflection and refraction in homogeneous nematic layers}
\label{sec:reflect}
The amplitude grating effect of a distorted nematic layer depends on the  amplitude $\mathbb{E}(\bm x, d)$ ( 
$\mathbb{O}(\bm x, d)$)  of the electric field component with  extraordinary (ordinary)  polarization  at the upper surface, $z =d$   (see Eq.~(\ref{eq:eik0})). The  amplitudes depend linearly on  their values  $\mathbb{E}_0$ ($\mathbb{O}_0$) at $z =0$. To make a full description of an experiment we have to connect the field amplitudes at the surfaces of the nematic layer to  
those of the adjacent glass plates. 

According to Eq.~(\ref{eq:wavedef}) the wavevector $\bm k = k_0 \bm k' = k_0(\bm k'_\parallel, k'_z)$ of the incident plane wave can be represented in terms of the polar angle
$\vartheta_g$ and the azimuthal angle $\phi$ of $\bm k'$ with respect to the normal ($\parallel
\bm{\hat z}$) of the nematic layer.  
As already indicated in Sec.~\ref{subsec:light} we have to distinguish between transverse electric (TE)
waves with the polarization vector $\bm p_E(\bm k')$ 
and the transverse magnetic ones (TM) with polarization vector  $\bm p_M(\bm k')$ where:
\be
\label{eq:polvec}
\bm p_E(\bm k') = \frac{\bm {\hat z} \times \bm k'} {|\bm {\hat z} \times \bm k'|}, \quad  \bm p_M(\bm k') = \frac{\bm p_E \times \bm k'} {|\bm p_E \times \bm k'|}.
\ee
In the case of perpendicular incidence ($\vartheta_g = 0$) we chose $\bm p_E =\bm{\hat y}$ and $\bm p_M =\bm{\hat x}$.

An  incident TE wave with amplitude $A$ and wavevector $\bm k' = (\bm k'_\parallel, k'_z >0)$ is partially reflected (wavevector $\bm k_r' = (\bm k'_\parallel, -k'_z)$) and partially transmitted at the glass-nematic interface. The parallel components of all participating wavevectors are the same,  while $|k'_z|$  is  determined by $|\bm k'_\parallel|$ and the respective refraction indices $n_{rf}$ in the different media as discussed in Sec.~\ref{subsec:light}. 
The general procedure to calculate  reflection and transmission of 
monochromatic plane waves at interfaces is discussed in many textbooks like \cite{Born}. A very clear presentation for nematics can be found in \cite{Igna12}.  

The continuity of the tangential electric field components leads at the glass-nematic interface $z =0$ to the condition:
\be
\label{eq:pE}
 A (\bm t \cdot \bm p_E(\bm k')) = A \,\bm t \cdot \Big ( R_{EE} \, \bm p_E(\bm k')  +  R_{EM} \, \bm p_M(\bm k_r') + T_{Ee}\, \bm {e} [\bm k_e'] + T_{Eo} \,\bm{o}[\bm k_o']\Big),  
\ee
both for $\bm t = \bm {\hat x}$ and $\bm t = \bm {\hat y}$, where the nematic polarization vectors $\bm e_0[\bm k_e'], \bm o_0[\bm k_o']$ are defined 
in Eq.~(\ref{eq:veco}).   
Note that according to Eq.~(\ref{eq:pE}) an incident TE wave 
will in general produce TE and TM reflected waves with amplitudes given by the reflection coefficients  $R_{EE}, R_{EM}$. In the nematic layer both ordinary and extraordinary contributions may exist with amplitudes determined by the transmission coefficients $T_{Eo}, T_{Ee}$. 
In analogy to Eq.~(\ref{eq:pE}) we have to exploit in addition the continuity of the tangential magnetic field components on the basis of Eqs.~(\ref{eq:maxs}).
It turns out that the necessary continuity of the normal components of the magnetic induction $\bm B = \mu \bm  H$ and of the 
dielectric displacement $\bm D$ is automatically fulfilled via the tangential conditions on $\bm E, \bm H$. 
Thus we have to solve four linear equations for the coefficients $R_{EE}, R_{EM}$ and 
and   $T_{Ee}, T_{Eo}$.
Alternatively, if the incident field is parallel to $\bm p_M$ we need the  reflection coefficients $R_{MM}$, $R_{ME}$ and the transmission coefficients $T_{Me}$, $T_{Mo}$, which are
defined in analogy to Eq.~(\ref{eq:pE}).  
At the nematic-glass interface ($z =d$) we have to distinguish between incident
ordinary plane waves ($\bm E \parallel \bm o_0$) and extraordinary ones  ($\bm E \parallel \bm e_0$) (see Eq.~(\ref{eq:veco})). Thus we  have to calculate 
the corresponding reflection- and  transmission coefficients $R_{ee}, R_{eo}, T_{eE}, T_{eM}$ and analogously $R_{oe}, R_{oo}, T_{oE}, T_{oM}$.  For instance if the incident electric field is parallel to  $\bm e_0$, the continuity condition for the tangential electric field reads  as follows: 
\be 
\label{eq:pole}
 A\,(\bm t \cdot  \bm e_0[\bm k'_e])  = A\ \bm t \cdot \Big [R_{ee} \,\bm e[({\bm k'_e})_r]  +  R_{eo}\, \bm o[({\bm k'_o})_r] + T_{eE}\, {\bm p_E(\bm k')} + T_{eM}\, \bm p_M(\bm k')\Big ].  
\ee

All reflection and transmission coefficients, which require only the solution of linear equations,  depend on 
the azimuth and polar angles  $\phi,\vartheta_g$ of  $\bm k'$ defined in 
Eq.~(\ref{eq:wavedef}) and on the refraction indices $n_g,n_o, n_e$. 
 The resulting general final expressions are, however, lengthy and not very transparent; thus they will not be reproduced here. In the special case of perpendicular incidence ($\vartheta_g =0$) we arrive at:  
\begin{align}
\label{eq:transcoef}
(T_{Eo},T_{Mo}) = ( \cos \phi, -\sin \phi)\frac{2 n_g}{n_g +n_o} \;, \quad
(T_{Me},T_{Ee}) = ( \cos \phi, \sin \phi)\frac{2 n_g}{n_g +n_e} \;, 
\nonumber\\ 
(T_{oE},T_{oM}) = ( \cos \phi, -\sin \phi)\frac{2 n_o}{n_g +n_o} \;, \quad
(T_{eM},T_{eE}) = ( \cos \phi, \sin \phi)\frac{2 n_e}{n_g +n_e} \;.
\end{align}
In experiments at most small $\vartheta_g$ are used.  
It then turns out that the transmission coefficients in Eqs.~(\ref{eq:transcoef}) remain unchanged to order $O(\vartheta_g)$ and that only the wavevectors and the polarization vectors $\bm p_M$, $\bm e_0$ acquire corrections $\propto \sin(\vartheta_g)$. 
If we concentrate in addition on incident TM waves with  $\phi =0$, only the transmission coefficients  $T_{Me}$, $T_{eM}$ come into play. In other words 
an incident TM wave leads only to a TM wave above the nematic layer.  

In this appendix we have concentrated on the electric field components of the electromagnetic field. Corresponding
reflection and transmission coefficients can be defined as well for the magnetic field
components. They are obtained for instance from the 
electric ones with the help of the Maxwell equations (\ref{eq:maxall}).

\section{Determination of the field amplitudes}
\label{sec:ampliE}

Here we sketch the derivation of the differential equations (see \cite{Pana}) which determine the field amplitudes $\mathbb{O}, \mathbb{E}$ and $\mathbb{Z}$ in Eq.~(\ref{eq:eik0}). 
First of all the terms of the order $O(k_0^2)$ from $ \bm V = \bm o^N$ and $ \bm V = \bm e^N$   vanish by construction of the eikonal equations.  Furthermore $\bm s^N$ does not give a contribution in this order since $\bm \nabla (S_e + S_0)/2 \parallel \bm s^N$.  

The terms $\propto  k_0$ from Eq.~(\ref{eq:iden}) can be represented  in term of the 
vector operator $\bm G\sembrack{\bm V|S} = (G_x, G_y, G_z)\sembrack{\bm V|S}$ defined as  
\be \label{eq:Gdef}
\begin{split}
G_i\sembrack{ \bm V|S} =  2(\partial_m S) (\partial_m V_i)  +  (\partial_m \partial_m S) V_i -
\partial_i  [(\partial_m S) V_m)] -(\partial_i S) (\partial_m V_m),\\ \hspace*{-1cm} \text{with} \quad \bm V = (V_x,V_y,V_z) \; \text{and}\;  i,m = (x,y,z).
\end{split}
\ee
 As usual, summation over the repeated indices $m$ is assumed.
Injecting the ansatz for $\bm E$ (Eq.~(\ref{eq:eik0})) into Eq.~(\ref{eq:el}) and collecting all terms $\propto k_0$ we arrive at:
\be \label {eq:Eoe}
 e^{\im k_0 S_o} \bm G\sembrack{ \mathbb{O} \bm o^N|S_o} +  e^{\im k_0 S_e} \bm G\sembrack{\mathbb{E}  \bm e^N|S_e} - e^{\im k_0 (S_e + S_o)/2} \mathbb{Z}\, \bm A = 0,
\quad \bm A := \bm \varepsilon\cdot  \bm s^N.
\ee
In line with \cite{Pana}  the amplitude $\mathbb{Z}$ is now eliminated by multiplying the term with $i =x $ in Eq.~(\ref{eq:Eoe})
with $A_z$ and subtracting the term with $i =z$   after a multiplication with $A_x$.
The analogous procedure is applied to the term with $i =y$. Thus we arrive at the final equations to determine the field amplitudes
$\mathbb{E}(\bm x,z), \mathbb{O}(\bm x,z)$:
\beal
\label{eq:ampE}
&A_z \big ( a_{-} G_x\sembrack{\mathbb{O} \bm o^N|S_o} + a_{+} G_x\sembrack{\mathbb{E} \bm e^N|S_e}\big) =
A_x \big ( a_{-} G_z\sembrack{\mathbb{O} \bm o^N|S_o} + a_{+} G_z\sembrack{\mathbb{E} \bm e^N|S_e}\big),
\nonumber\\
&A_z \big( a_{-} G_y\sembrack{\mathbb{O} \bm o^N|S_o} + a_{+} G_y \sembrack{\mathbb{E} \bm e^N|S_e} \big) =
A_y \big( a_{-} G_z\sembrack{\mathbb{O} \bm o^N|S_o} + a_{+} G_z\sembrack{\mathbb{E}\bm e^N|S_e}\big),\nonumber\\
&\text{where} \quad a_\pm = \exp[\pm ik_0 (S_e -S_0)/2].
\end{align}

\section{Magnetic field in order $O(\theta_m^2)$}
\label{sec:mag2}
Expanding Eq.~(\ref{eq:eikob}) to second order in $\theta_m^2$ yields the following equation for the expansion coefficient $B^{(2)}(x,z)$ defined in Eq.~(\ref{eq:Bexpand}): 
\be
\label{eq:b2h}
\begin{split}
2 n_e \frac{\partial}{\partial z} B^{(2)}(x,z) &= -2 \beta  n_e \delns_z(x,z) \left [
B_0 \frac{\partial}{\partial z} \delns_z(x,z) + \frac{\partial}{\partial x} B^{(1)}(x,z) \right ]\\
&- n_e \beta B^{(1)}(x,z)
\frac{\partial}{\partial x} \delns_z(x,z)
-B_0 \left[ \frac{n_e^2}{n_o^2} \partial_{xx} S^{(2)}(x,z) + \partial_{zz} S^{(2)}(x,z) 
\right] ,
\end{split}
\ee
where $B^{(1)}(x, z)$ (Eq.~(\ref{eq:bsolveone})) and $S^{(2)}(x,z)$ 
(see Eq.~(\ref{eq:S2ord}))  have been already calculated.
After  straightforward integration over $z$ we obtain:
\be \label{eq:B2fin}
B^{(2)}(x,d) = -B_0 \, \frac{q^2 d^2\beta}{8 \pi^2} {\Big (} 2 \beta
+[ 6 \beta + \pi^2 (\beta +1)\,] \cos(2 q x) {\Big )}.
\ee
Furthermore, by collecting the terms $\propto \theta_m|\bm k'_\parallel|$ in Eq.~(\ref{eq:eikob}) 
we obtain 
the following equation for $B_y^{(1,1)}$ in Eq.~(\ref{eq:Bexpand}):
\begin{align} \LB{eq:Bfull}
2 n_e \frac{\partial}{\partial z} B^{(1,1)}(x,z)&=
- B_0 n_e \beta \frac{\partial}{\partial x} \delns_z  -B_0  k'_x \beta \frac{\partial \delns_z}{\partial z}\\ &-2 \frac{k'_x n_e^2}{n_o^2} \frac{\partial B^{(1)}(x,z)}{\partial x}
- B_0 \frac{n_e^2}{n_o^2} \frac{\partial^2
S^{(1,1)}}{\partial x^2} - B_0  \frac{\partial^2
S^{(1,1)}}{\partial z^2}\nonumber.
\end{align}
By performing the $z$-integration within the one-mode approximation we arrive at:
\be \LB{eq:byfinal}
B^{(1,1)}(x,d) = B_0 \beta \dfrac{n_e}{n_o^2 \pi} d^2 q^2
\cos(q x).
\ee

\section{The nematic layer as a diffraction grating}
\label{sec:talbot}
The electric field of the wave propagating through the nematic layer from  $z =0$ to $z =d$ is in  general described by Eq.~(\ref{eq:eik0}); finally we need the field at the upper surface of the layer ($z = d$).  At first we may safely neglect the $O(k_0^{-1})$ contribution $\propto \mathbb{Z}$. In any case  since $\bm s^N \approx \bm{\hat  z}$ for small $\vartheta_g$ the associated  plane waves 
propagate practically in the $xy$-plane and are thus not  relevant in the present context.  
With respect to the extraordinary and the ordinary field components 
their  polarization vectors simplify considerably   
at $z =d$.  
We have $\bm o^N \rightarrow  \bm o_0[\bm k'_o]$ , $\bm e^N \rightarrow \bm e_0[\bm k'_e]$, where the constant vectors $\bm o_0, \bm e_0$ are defined in Eq.~(\ref{eq:veco}) with 
$\nv = \bm {\hat x}$. The corresponding amplitudes $\mathbb{O},\mathbb{E}$, which are periodic 
in the horizontal coordinates $\bm x =(x,y)$, describe obviously the 
amplitude grating effect of the nematic layer.  In view of the general decomposition $S_e(\bm x,z) = S_e^0( \bm x,z) + \bar S_e(\bm x, z)$ (see Eqs.~(\ref{eq:Sbas}),  (\ref{eq:Sexpan})), the phase grating effect of the nematic layer is captured by $\bar S_e(\bm x, d)$. This quantity  is periodic in $\bm x$ as well due to the sums of trigonometric functions 
with the arguments $\bm q\cdot \bm x$ and  $2 \bm q\cdot \bm x$ originating from the solutions of the eikonal 
equation Eq.~(\ref{eq:eikz}) to order $O(\theta_m^2)$. For simplicity we concentrate on the discussion of  
the crucial extraordinary contribution $E_e(\bm x)\bm e_0[\bm k'_e]$ with $E_e (\bm x) = \mathbb{E}(\bm x,d) \exp[\im S_e(\bm x,d)]$.  The total electric field  $\bm E_e(\bm x,d)$ can be written as a Fourier series 
as follows:
\be
\label{eq:fieldn1}
\bm E_e(\bm x, d) = \mathbb{E}_0 \exp{[\im k_0 (\bm k'_\parallel \cdot \bm x + {k'_z}^e(\bm k') d)]} \sum_{n = -\infty}^{\infty} C^N(n) e^{i\,n \scal{q}{x}}\,
\bm e_0[n \bm q/{k_0} + \bm k'],
\ee
where the phase  prefactor comes from $S_e^0$ (Eq.~(\ref{eq:Sbas})) evaluated at $z = d$.  
The Fourier series contains the product of the corresponding ones from $\mathbb{E}(\bm x, d)$
and from  the phase factor $\exp[\im k_0 \bar S_e(\bm x, d)]$. The latter  can be transformed
into a Fourier series by exploiting   
the identities
 \be
\label{eq:besselexp}
e^{\im \alpha \sin \beta} = \sum_{n = -\infty}^{\infty} J_n(\alpha) e^{\im n \beta},
\; e^{\im  \alpha \cos \beta} = \sum_{n = -\infty}^{\infty} i^n J_n(\alpha) e^{\im  n \beta}, 
\ee
where $J_n(\alpha)$ denotes the Bessel function of the first kind. For small $\theta_m$, 
on which we will mainly concentrate,  it is sufficient to truncate the expansion of $\mathbb{E}(\bm x, d)$  at $\theta_m$, since in general $k_0\bar S_e(\bm x,d)$  prevails. To obtain closed analytical expressions   we use the one-mode approximation for $\deln$ (Eq.~(\ref{eq:ansatz_n})) to evaluate the terms  given in Eqs.~(\ref{eq:efieldfine}), (\ref{eq:Setot}) and arrive at:
\beal
\label{eq:expcoef}
&\mathbb{E}(\bm x, d) = \mathbb{E}_0 ( 1 + c_{E1} \theta_m \sin (\bm q \cdot \bm x )) , c_{E1} = (q_x d) \frac{\beta}{\pi} \;,
\nonumber\\
&k_0 \bar S(\bm x, d) = \cos (\bm q \cdot \bm x) (\theta_m c_{S1} + \theta_m^2 c_{S2} \cos (\bm q \cdot \bm x)) \;\text{with}
\nonumber\\ 
&c_{S1} = -(k_0 d) \frac{2 \beta}{\pi} \cos(\phi) \sin(\vartheta_g) \;, \;\;
c_{S2} =   -(k_0 d) n_e \frac{\beta}{4} .
\end{align}
Expanding thus the product $\mathbb{E}(\bm x, d) \exp[\im k_0 \bar S_e(\bm x, d)]$ we obtain the following approximations for the  expansion coefficients $C^N(n)$ in Eq.~(\ref{eq:fieldn1}):
\beal
\label{eq:Ccoef}
C^N(0) & = 1 + \theta_m^2 \frac{1}{4}[2 \im c_{S2} - c_{S1}^2] \;, \;\;
C^N (\pm 1) = \im  \theta_m  \frac{1}{2}
(c_{S1} \mp c_{E1}), \nonumber\\  
&C^N(\pm 2) =  
\im \theta_m^2 \frac{1}{8}( 2 c_{S2} + i c_{S1}^2 \pm \im 2 c_{S1} c_{E1}).   
\end{align}
%
Since the electric field (Eq.~(\ref{eq:fieldn})) at $z =d$ is expressed as a superposition of plane waves it is easy to construct the electric field in the adjacent glass plate of thickness $d_g$.  Here we have in general 
a superposition of TM and TE waves. Let us start with the TM waves which have the 
general representation:
\beal
 \label{eq:wavg}
\bm E_M^G (\bm x,z) &= \mathbb{E}_0  \exp  [i k_0 \bm k'_\parallel \cdot \bm x] \sum_{n=-\infty}^{\infty} C^G(n)  \exp  [i( n \bm q \cdot \bm x + z k^G_z(n))]\bm p_M(\bm k' + n \bm q/{k_0}) \;, \nonumber\\ & \text{for} \;  d \le z \le d + d_g
\; \text{with} \; k^G_z(n) = k_0\sqrt{ n_g^2  -(\bm k_\parallel' + n \bm q/{k_0})^2} \;.
\end{align}
We are only interested in the propagating waves where the argument of the square root 
in Eq.~(\ref{eq:wavg}) is positive, which restricts the summation over $n$ to a cutoff $n =
n_{cut}$. For the typical nematic  pattern $n_{cut} \gg 1$ holds. The contributions for $|n| > n_{cut}$ which decay exponentially with increasing 
$z$, i.e. the ``evanescent'' waves, are not recorded in the standard experiments.
In fact, only the Fourier coefficients for small $|n| < 3$ will play an important role for small 
distortion amplitudes $\theta_m$. Thus besides  the leading terms in $|\bm k'_\parallel| =\sin(\vartheta_g)$ only  the leading terms in 
the small quantity $|n \bm q|/k_0$ are kept in the following. 
For instance the transmission coefficient $T_{ga}$ from a glass to an air layer with refraction index $n_a =1$ is given 
in this approximation as \cite{Born}:
\be
\label{eq:transga}
T_{ga}(\bm k' + n \bm q/k_0)  = \frac{2 n_g}{n_g + n_a}\,.
\ee
Matching to the electric field (Eq.~(\ref{eq:fieldn})) in the nematic layer at $z =d$
yields: 
\be
\label{eq:kzg}
C^G(n) \exp[\im  k^G_z(n) d] = T_{eM}(\bm k')  \exp[\im k_0 {k'_z}^e(n) d]\, C^N(n).
\ee
The wave gets then refracted again at the glas-air interface ($z =d+d_g$) and propagates 
further in air. 
The electric field in air for $z \ge d + d_g$ has the same representation as in 
Eq.~(\ref{eq:wavg}) with Fourier coefficients $C^A(n)$ instead of  $C^G(n)$.  Furthermore
we need  $k^A_z(n)$  where $n_g$ in $k^G_z(n)$ (Eq.~(\ref{eq:wavg})) 
is replaced by $n_a =1$. Matching the electric fields  at $z = d + d_g$ yields:   
\be
\label{eq:kzga}
C^A(n) \exp[\im  k^A_z(n) (d +d_g)] = T_{ga} (\bm k' + n \bm q/k_0)  \exp[\im k_0 k^G_z(n) (d+d_g)] C^G(n),
\ee 
where $T_{ga}$ is given in Eq.~(\ref{eq:transga}).
At the end we arrive at:
\beal 
\label{eq:Bair1}
\bm E^A (\bm x,z) &= \mathbb{E}_0 \exp[\im k_0 (\bm k'_\parallel \cdot \bm x + z')]\mathfrak{S}(\bm x, z') 
\quad \text{with} \; z' = z-(d+d_g) \; \text{and}
\\&\hspace{-1.5cm}\mathfrak{S} = \sum_{n=-n_{cut} }^{n_{cut}} C_n^N 
T (d_g,n) \exp  [\im  (n \bm q \cdot \bm x) + \im z' (k^A_z(n) - k^A_z(0))]\bm p_M(\bm k' + n \bm q/{k_0}) 
\end{align}
The transfer function $T(d_{g},n)$ describes the effect of the glass layer where 
\be 
\label{eq:deftrans}
T(d_{g},n)= T_{eM}(\bm k' + n \bm q/k_0)  T_{ga} (\bm k' + n \bm q/k_0) \exp[i d_{g} k^G_z(n) + i k^e_z d].
\ee 

The quantitative evaluation of the electric field in Eq.~(\ref{eq:Bair1}) is straightforward but requires numerical effort. We concentrate here on
analytical expressions which describe all relevant features for small
$\theta_m$ very well.  
According to Eq.~(\ref{eq:Ccoef}) only the Fourier coefficients for $|n| < 3$ come into play. Exploiting in addition the smallness of $\vartheta_g$ and of $|n \bm q|/k_0$
all transmission coefficients like $T_{ga}$ (see Eq.~(\ref{eq:transga})) can be safely replaced by their
values for  perpendicular incidence and can be taken out from the sums. An analogous approximation   applies also to the polarization vectors. 
In the exponents of  $\exp[\im k_0 (k^A_z(n) - k^A_z(0))z']$ we keep in the spirit of the paraxial approximation  the leading term by expanding
inside the square roots with respect to the small quantity $q^2/k_0^2$.  
Thus starting from Eq.~(\ref{eq:Bair1}) the electric field is well approximated by:
\beal 
\label{eq:represen}
\bm E^A(\bm x, z) &= \mathbb{E}_0 T_{eM} T_{ga} \exp[\im k_0 (\bm k'_\parallel\cdot \bm x + z')] \;  \mathfrak{S}'(\bm x, z')\,\bm p_M(\bm k'),  \\
\mathfrak{S}' &= 
 C^N(0) 
  +\big(C^N(+1)  \exp[\im \bm q \cdot \bm x]  + C^N(-1) \exp[-\im \bm q \cdot \bm x]\big) \exp[-\im \frac{q^2}{2 k_0} z']\nonumber\\
   &+\big(C^N(+2)  \exp[\im \bm 2 q \cdot \bm x] + C^N(-2) \exp[-\im 2 \bm q \cdot \bm x]\big) \exp[-\im \frac{4 q^2}{2 k_0} z'] + \cdots
\end{align}
with $z' = z -(d+d_g)$. 
 Hence from the knowledge of the $C^N(n)$  we obtain immediately
the relative intensity of the refraction fringes $I_n = |C^N(n)|^2$;  
the shadowgraph intensity  $I_S(\bm x,z)$ is determined by  $|\bm E^A (\bm x,z)|^2$.
Further details will be discussed in Sec.~\ref{sec:discuss}.

We have given explicit results for TM waves entering the nematic layer at $z=0$ and leaving at $z =d$.  
In this case the coupling to the extraordinary waves in the nematic 
layer, which are associated with strong phase grating effects,  is guaranteed. This is the situation
used in the typical experimental setups. The analysis of other cases, however, would  follow step by step the same calculational scheme, We had to use the ordinary field
amplitude $\mathbb{O}$ in the nematic layer which would 
then couple to the polarization vectors $\bm p_E$ in
the glass plates.

\section{Discussion of the eikonal solutions}
\label{sec:eikonaln}

In Sec.~\ref{sec:discuss} we have demonstrated 
that an inclination of the wavevector $\bm k'$ of the incident light even by a small polar angle $\vartheta_g$ with respect to the $\bm{\hat z}$ (see  Eq.~(\ref{eq:wavedef})) may have a strong effect:  both the intensity of the first order ($n =1$) diffraction fringes and the contrast of the shadowgraph pictures considerably increase. 
So far  we have concentrated on  the terms linear 
in $\vartheta_g$ and to zero azimuthal angle $\phi$. This section is devoted to the question, 
whether the use of a rotation of the incidence plane about the $z$-axis, i.e. finite  $\phi$,  will give additional advantages.   
For that purpose we have performed  a systematic expansion of $\bar  S_e$ (see Eq.~(\ref{eq:eikz}))   up to second order in the director amplitude $\theta_m$  by using the following ansatz:
\be
\label{eq:eik2ord}
\bar S_e(\bm x, z) =  \theta_m S_e^{(1)}(\bm x, z)  + \theta_m^2 S_e^{(2)}(\bm x, z) \;.
\ee

The expressions for the coefficients $A$, $B$ (see Eq.~(\ref{eq:eikfinAB}))  to be used in Eq.~(\ref{eq:eikzwur}) have not to be modified, but   $C$ has to be generalized as follows:
\be
\label{eq:eikfinC}
\begin{split}
C &= \theta_n (C^1_{\delns} + C^1_S) + \theta^2_n (C^2_{\delns} + C^2_S) \quad \text{with}\\ 
C^1_{\delns} &= - 2 \beta k'_x ({k'_z}^e \delns_z + k'_y \delns_y) \;, \;\; C^1_S = - (k'_y \frac{\partial}{\partial y} \bar S_e + (n_e/n_o)^2 k'_x \frac{\partial}{\partial x} \bar S_e) \;, \\
C^2_{\delns} &= 
\beta \big( -2 k'_y {k'_z}^e \delns_z \delns_y +({k'_x}^2 - {k'_y}^2) 
\delns_y^2 +  
({k'_x})^2 - ({k'_z}^e)^2) \delns_z^2\Big ) \;, \\
C^2_S &=   -[2 \beta k'_x \delns_y  +\frac{n_e^2}{n_o^2} \frac{\partial}{\partial y} \bar S_e]\frac{\partial}{\partial y} \bar S_e 
- [\frac{\partial}{\partial x} \bar S_e + 2 \beta (k'_y \delns_y + {k'_z}^e \delns_z)  ]\frac{\partial}{\partial x} \bar S_e, 
\end{split}
\ee
where ${k'_z}^e$ has been defined in Eq.~(\ref{eq:kz}).

To expand the eikonal solution of Eq.~(\ref{eq:eikzwur}) to order $O(\theta_m^2)$ we need 
also the following relation:
\be 
\label{eq:expanen}
\Big (B + \sqrt{B^2 + AC} \Big )^{-1} =  
\frac{1}{2 {k'_z}^e} 
\Big ( 1 - \theta_m  
  \frac{4 {k'_z}^e  \beta k'_x (k'_y \delns_y 
- {k'_z}^e \delns_z)
 + k'_y \frac{\partial}{\partial y} \bar S_e - (n_e/n_o)^2 \frac{\partial}{\partial x} \bar S_e } {4  ({k'_z}^e)^2} \Big).    
\ee
In order $\theta_m$ we obviously arrive from Eq.~(\ref{eq:eikzwur}) at the following differential equation for $S_e^{(1)}$:
\be
\label{eq:linS}
\frac{\partial}{\partial z}  S^{(1)}_e - \frac{1}{2 {k'_z}^e} S^{(1)}_e  - \big(k'_y \frac{\partial}{\partial y} S^{(1)}_e + (n_e/n_o)^2 k'_x \frac{\partial}{\partial x}  S^{(1)}_e\big) = inh(\bm x, z) \equiv  C^1_{\delns} \;.
\ee
Let us now switch to Fourier space: 
\be
\label{eq:switchfou}
 S^{(1)}_e(\bm x, z) = \tilde S^{(1)}_e(\bm q, z) \exp[\im \bm q \cdot \bm x] +c.c. \;, \;\;
inh(\bm x, z) = \widetilde{inh}(\bm q,z) \exp[\im \bm q \cdot \bm x] +c.c.  
\ee
where $\partial_{\bm x}  \rightarrow i \bm q$ in Eq.~(\ref{eq:linS}).
Thus 
the solution of Eq.~(\ref{eq:linS}) in Fourier space with initial condition $\tilde  S^1_e(\bm q, z) =0$ for $z =0$ is easily obtained as:
\beal
\label{eq:Ssolv}
 \tilde S^1_e(\bm q, z) &= \exp[-\im \lambda(\bm k',\bm q) z/d ]
 \int_0^z dz' \exp[\im \lambda(\bm k',\bm q )  (z'/d)]\, \widetilde{inh}(\bm q, z'),\nonumber\\
& \text{with}  \quad \lambda(\bm k', \bm q) = -\frac{k'_y q_y + (n_e/n_o)^2 k'_x q_x}{{k'_z}^e}.
\end{align}
Returning to position space we arrive from Eq.~(\ref {eq:switchfou}) at the following 
general representation of $S^{(1)}_e(\bm x, z)$:
\be
\label{eq:S1}
 S^{(1)}_e(\bm x, z) = \sin(\bm q \cdot \bm x) f_s(z) +  \cos(\bm q \cdot \bm x) f_c(z).     
\ee 
The functions $f_s$, $f_c$ are obtained by performing the $z$-integration in Eq.~(\ref{eq:Ssolv}).  Their specific form depends on the ansatz chosen for $\delns_y$, $\delns_z$ which according to Eq.~(\ref{eq:linS}) determine $inh(\bm x, z)$. 
The phase-grating effect to order $\theta_m$ is determined by $S^{(1)}_e(\bm x, z =d)$; an explicit analytical expression is obtained again within the  
one-mode approximation for $\deln$ (Eq.~(\ref{eq:ansatz_n})). 
The lengthy expressions simplify considerably if we 
confine ourselves in addition to the leading terms in $|\bm k'_\parallel| \propto \sin (\vartheta_g)$ according to Eq.~(\ref{eq:wavedef}), which give already the main insight into the relevance of the various 
contributions to  $ S^{(1)}_e(\bm x, z =d)$.  Restricting ourselves 
to the terms up to order $O(\sin^2(\vartheta_g))$ leads to:
\beal 
\label{eq:fcsexpa}
f_s(d) & = -d \beta \frac{\sin(\vartheta_g)^2 \cos \phi}{n_o^2 \pi} \big ( 2 n_o^2 a_y \sin \phi  +  n_e^2 \cos \phi   q_x + n_o^2 \sin \phi q_y\big) + O(\sin^4(\vartheta_g)) \;, \\
\label{eq:fcsexpb}
f_c(d) &= -d \frac{2  \beta}{\pi} \sin(\vartheta_g) \cos \phi + O(\sin^3(\vartheta_g)).
\end{align}
Note that  in the  
case of perpendicular incidence ($k'_x, k'_y =0)$ we get no contribution to phase grating of the order $O(\theta_m)$;  the only linear contribution $\propto 
 k'_x \cos(\bm q \cdot \bm x)$, which we have obtained already before in Eq.~(\ref{eq:S2ord}), requires a nonzero $\vartheta_g$. 
We have to keep the terms $\propto \sin^2(\vartheta_g)$ in order to identify a contribution 
of the director twist ( $\sim \delns_y$) and of an in-plane rotation of $\bm q$ (finite $q_y$).

Turning to quadratic order in $\theta_m^2$ we have to solve an equation for the expansion coefficient $\bar S^{(2)}_e$ (see Eq.~(\ref{eq:eik2ord})), which has the same structure 
as Eq.~(\ref{eq:linS}) except a different inhomogeneity $inh_2(\bm x, z)$.  
Here we have contributions from $C^2_{\delns}$
and from $C^2_{S}$ in Eq.~(\ref{eq:eikfinC}) where the solution $\bar S^{(1)}_e$ given in Eq.~(\ref{eq:S1}) has to be used. In addition  
we find contributions
from the product of the term $\propto \theta_m$ in Eq.~(\ref{eq:expanen}) and the terms 
$(C^1_{\delns} +   C^1_{S})$ in  Eq.~(\ref{eq:eikfinC}). In analogy to the treatment of Eq.~(\ref{eq:linS}) and its solution shown in Eq.~(\ref{eq:fcsexpa})  we arrive at the following general representation for 
$\ S_e^{(2)}(\bm x, z)$:
\be
\label{eq:S2}
S^{(2)}_e(\bm x, z) =  f_0^{(2)}(z) + \sin(2 \bm q \cdot \bm x) f^{(2)}_s(z) + \cos(2 \bm q \cdot \bm x) f^2_c(z).
\ee 
Performing the required $z$-integrations within the one-mode approximation and restricting ourselves 
to the terms up to order $O(\sin(\vartheta_g))$ the analytical expressions read as follows:
\bsub
\label{eq:thetm2}
\beal
& f_0^{(2)}(d) +  f_c^{(2)}(d) \cos(2 \bm q \cdot \bm x)   =  - d \frac{n_e \beta}{8} (1 + \cos(2\bm q \cdot \bm x)) + O(\sin^2(\vartheta_g), \label{eq:cos2}\\ 
&f_s^{(2)}(d) = -d \beta \frac{\sin(\vartheta_g)}{8 }\Big ( a_y \sin \phi + 4 (\beta/{\pi^2} + \frac{n_e^2}{n_o^2}) q_x d  \cos \phi +   q_y d \sin \phi \Big ) +  O(\sin^3(\vartheta_g)).
\end{align} 
\esub
 
As to be expected  the $\vartheta_g$-independent contribution to $S_e^{(2)} (\bm x, d)$ in Eq.~(\ref{eq:cos2})  is equal  to the one already derived in Eq.~(\ref{eq:Setot}).  The new term  $f_s^{(2)}(d) \propto \sin(\vartheta_g)$ reveals the impact of a director twist ($\sim \delns_y$) and of an in-plane rotation of $\bm q$ (finite $q_y$).

\section{Shadowgraphy in optically isotropic media: Rayleigh-B\'enard convection}
\label{sec:scalopt}

In the following we comment briefly on the short-wavelength expansion technique  for Rayleigh-B\'enard convection (RBC), where we follow closely the notations in \cite{cannell}.
The convection cell has the thickness $d$ ($0< z < d$) with $T_1 > T_2$ the prescribed
temperatures at the lower and upper plate, respectively.
In the convective state the temperature distribution is given as:
\be \label{eq:temp}
T(x,y,z) = T_0 - \Delta T \frac{z-d/2}{d}+ \Theta_{conv}(x,y,z),\; \text{with}\;
T_0 = \frac{T_1 + T_2}{2}, \,\Delta T = (T_1 -T_2),
\ee
where $\Theta_{conv}(x,y,z)$ denotes the convective temperature contribution, which is available as a Galerkin expansion from standard codes.  
We consider the fluid as an isotropic medium with a space-dependent dielectric
permeability  $\varepsilon(x,y,z)$ and constant magnetic permeability 
$\mu =1$. Thus the refraction index is given as $n^2 = \varepsilon$.
It  depends on the density $\rho$, which varies with temperature in the RBC case. Thus we use an expansion about the mean temperature $T_0$:
\be \label{eq:nB}
n(\rho(T)) = n(\rho(T_0)) + \frac{\partial n}{\partial \rho} \frac{d \rho}{d T}\Big|_{T = T_0} (-\Delta T \frac{z-d/2}{d} + \Theta_{conv}(x,y,z)).
\ee
According to Eq.~(1) in Ref.~\cite{cannell}, 
the three terms on the rhs of Eq.~(\ref{eq:nB}) are parametrized as: 
\be \label{eq:nrefract}
n(x,y,z) = n_0 + n_{heat} + n_{conv} \equiv  n_0 + n_2 z/d + n_1 \sum_{i=1}^N a_i(x,y) b_i(z),
\ee
where the coefficients $n_0$, $n_2$ describe the heat conduction state, while the term $n_{conv} \propto n_1$
measures the overall amplitude of $\Theta_{conv}$ in an Galerkin expansion. 

The starting point  for the optical analysis of RBC patterns  are the following wave equations for the electric
field, $\bm E$, and the magnetic field, $\bm H$,  with a monochromatic time dependence (see, e.g., Eqs.~(5, 6) in
\cite{Born}):
\bea \label{eq:wave}
\Delta \bm E &+& n^2 k_0^2 \bm E+ 2 \grad \left (\bm E \cdot   \bm \nabla (\ln\, n \right)= 0,\label{eq:wavel}\nonumber\\
\Delta \bm H &+& n^2 k_0^2 \bm H + 2 [\bm \nabla (\ln\, n )] \wedge \rot \bm H  = 0.\label{eq:wavmag}
\eea
The short-wave expansion is based on the ansatz:
\be \label{eq:shortwave} 
\bm  E = \bm e(\bm r)  \exp[i k_0 S(\bm r)] \;, \;\; \bm  H = \bm h(\bm r) \exp[i k_0 S(\bm r)] \;, \;\; \bm r = (x,y,z).
\ee
It is easy to see that we arrive from Eqs.~(\ref{eq:wavel}) at the
following equations for $\bm e(\bm r)$ (see \cite{Born}, Eqs.~(16)):
\bal
\label{eq:eikampe}
 &(n^2 - (\bm \nabla S)^2)\bm e  - \frac{1}{i k_0} \bm L^{e} (\bm e,S, n)= 0 + O(k_0^{-2})  \quad \text{with} \nonumber \\
 &\bm L^{e}( \bm e, S,n) = (\Delta S\, \bm e + 2 \left[\bm e \cdot \grad (\ln n)\right]) \grad S +  2 (\grad S \cdot \bm \nabla) \cdot \bm e. 
\end{align}
Analogously we obtain:
\bal
\label{eq:eikamph}
&(n^2 - (\bm \nabla S)^2)\bm h - \frac{1}{i k_0} \bm L^{h} (\bm e,S, n)= 0 + O(k_0^{-2}) \quad \text{with} \nonumber\\
&\bm L^{\bm h}(\bm h, S, n) = 
\left[2 \grad S \cdot \grad (\ln n) -\Delta S \right] \,\bm h   
- 2 \left[\bm h \cdot \grad (\ln n)\right] 
\grad S - 2 [\grad S \cdot \bm \nabla] \bm h, 
\end{align}

The leading order terms in Eqs.~(\ref{eq:eikampe}), (\ref{eq:eikamph}) yield the eikonal  equation:
\be \label{eq:eikonrbc}
(\bm \nabla S(x.y,z))^2 = n^2 (x,y,z),
\ee
while the next order terms ($\propto k_0^{-1}$) determine the amplitudes
$\bm e, \bm h$, respectively (see the remarks in \cite{Born} before  Eqs.~(41),(42) there).

We will here only address a 2D configuration with perpendicularly incident light in analogy to the 
planar case in nematics. The refraction index 
(Eq.~(\ref {eq:nrefract})) varies thus only in the $xz$-plane (''convection
rolls''), which is also the incidence plane of the light with $\bm H$
in the $y$-direction.
In the heat conduction state ($a_i \equiv 0$ in Eq.~(\ref{eq:nrefract})) we obtain immediately:
\be \label{eq:Ssolve0} 
S\equiv S^{(0)}(z) = n_0 z+ n_2 \frac{z^2}{2 d}.  
\ee
Using the ansatz: 
\be
\label{eq:defSrbc}
S = S^0 + S' \equiv S^0 + n_1 S^{(1)} + n_1^2 S^{(2)} + \cdots 
\ee
and Eq.~(\ref{eq:nrefract}) for the 
refraction index, we obtain from Eq.~(\ref{eq:eikonrbc})
\be \label{eq:eikrbc1}
2(n_0 + n_2 z) \frac{\partial}{\partial z} S'(x,z) + (\frac{\partial}{\partial x} S')^2 = 
2 (n_0 + n_2 z) n_{conv}(x,z) + n^2_{conv}(x,z). 
\ee
In linear order in $n_1$, $n_2$ Eq.~(\ref{eq:eikrbc1}) can be directly 
solved and we arrive at:
\be \label{eq:Ssolve1} 
S^{(1)}(x,z) = \sum_{i=1}^N \frac{\partial}{\partial x} a_i(x) \int_0^z dz' b_i(z').
\ee
Thus in contrast to the nematic case we obtain already phase modulation
in first order in $n_1$.
For the solution $S^{(2)}$ proportional to $n_1^2$ (see Eq.~(\ref{eq:defSrbc})) we obtain  
from Eq.~(\ref{eq:eikrbc1}) the following ODE in $z$: 
\be
\label{eq:eikonrbc2}
n_1^2 \left( 2(n_0 + n_2 z) \frac{\partial}{\partial z} S^{(2)}(x,z) + [\frac{\partial}{\partial x} S^{(1)}(x,z)]^2 + [\frac{\partial}{\partial z} S^{(1)}(x,z)]^2 \right)
= n_{conv}^2(x,z) 
\ee
which can be solved by a simple $z$-integration. Note that the term  $n_{conv}^2$  on the rhs
of Eq.~(\ref{eq:eikonrbc2}) cancels against $[\partial_z S^{(1)}]^2$. 
Thus we arrive at: 
\be \label{eq:Ssolve2}
S^{(2)}(x,d) = - \frac{1}{2 n_0} \int_0^d dz \sum_{i,j} \frac{d a_i(x)}{d x} \frac{d a_j(x)}{d x}
\int_0^z \int_0^z dz' dz'' b_i(z') b_j(z').
\ee
Consequently one obtains the following expression for the total phase modulation term at $z = d$:
\begin{equation}
 k_0 S(x,y,d) = k_0 S^{(0)}  + k_0 n_1 S^{(1)}(x,d)
+ k_0 n_1^2  S^{(2)}(x,d). 
\end{equation}
When we use the expression for  $S^{(0)}, S^{(1)}, S^{(2)}$ given in 
Eqs.~((\ref{eq:Ssolve0}), (\ref{eq:Ssolve1}),  (\ref{eq:Ssolve2})) we agree with Ref.~\cite{cannell} up to order $n_1$ but
disagree in order $n_1^2$, though the corresponding terms look very similar.

Since $\bm h$ has only a nonzero $y$-component, $h_y$, in the present geometry, it is convenient
to determine the amplitude modulation contribution 
from Eq.~(\ref{eq:eikamph}).  To order $k_0^{-1}$ , starting from Eq.~(\ref{eq:eikamph}), we  have to solve $\bm L^{\bm h} h_y(x,z) = 0$ with the boundary condition $h_y(x,0) = H_0$
where $H_0$ denotes the magnetic field amplitude of the incident plane wave. 
Explicitly written down,  $\bm L^{\bm h} h_y = 0$ reads as follows:
\be \label{eq:amphfin} 
\Big(2 \left[\frac{\partial}{\partial x} S \,\frac{\partial}{\partial x} + \frac{\partial}{\partial z} S \, \frac{\partial}{\partial z} \right] \ln (n(x,z)) 
- \Delta S \Big) h_y(x,z)
 - 2 \left[\frac{\partial}{\partial x} S \, \frac{\partial}{\partial x} + \frac{\partial}{\partial z} S \, \frac{\partial}{\partial z}\right] h_y(x,z) = 0.
\ee                                            
Note that the term $\grad S[\bm h \cdot \grad \ln n]$ in  Eq.~(\ref{eq:eikamph}) has vanished
identically.
 Equation (\ref{eq:amphfin}) is solved iteratively by using
the ansatz:
\be \label{eq:ansatzh}
 h_y(x,z) = h^0_y(z) + n_1 h^{(1)}_y(x,z) + O(n_1^2) .
\ee
In the heat conducting state (no $x$-dependence) Eq.~(\ref{eq:amphfin}) simplifies 
to:
\be \label{eq:amphfin1} 
\Big(\frac{\partial}{\partial z} S \, \frac{\partial}{\partial z}  \ln (n_0 z+ n_2 z^2/d) 
- \partial_{zz}S \Big) h_y(z)
 -  2 \frac{\partial}{\partial z} S \, \frac{\partial}{\partial z}  h_y(x,z) = 0.
\ee 
where only $S \equiv S_0$ from Eq.~(\ref{eq:Ssolve0}) has to be used.
Neglecting the small $n_2$ contributions (they can easily be incorporated) the incident
amplitude is not modified, i.e $h^0_y(z) \equiv  H_0$.
In order $n_1$ we obtain the ODE:
\begin{equation}
2 n_0 \frac{\partial}{\partial z} h^{(1)}(x,z) = -\partial_{xx} S^{(1)}(x,z) 
\end{equation}
which leads  to the following magnetic field amplitude:
\begin{equation}
h_y(x,z) = H_0 \left (1 - \frac{n_1}{2 n_0} \sum_{i=1}^N (\frac{\partial^2}{\partial x^2} a_i(x)) \int_0^z dz' b_i(z') \right).
\end{equation}
This expression agrees perfectly with Eq.~(20) in Ref.~\cite{cannell}.
Thus it has been proven, that also in RBC our calculational scheme needs only a few systematic  steps
to reproduce the previous results in \cite{cannell}, which were obtained after tedious
calculations. 
It is obvious, that oblique incidence can be treated without difficulty within our calculational scheme as well.
Also proceeding to higher order terms in $n_1$ is straightforward.

\begin{acknowledgments}
We are grateful to N. \'Eber for carefully reading the manuscript and useful remarks.
\end{acknowledgments}



\begin{thebibliography}{10}
\bibitem{problem}
For general deficiencies of geometric optics see, for instance, the introduction of \cite{cannell}. In section VI of this paper a minor technical error in \cite{rasenat89,rehberg89} is mentioned and also an unphysical divergence in second  order in the pattern amplitude there.
\bibitem{can_compare} 
The expression for $S^{(1)}$, $S^{(2)}$ are given in Eq.~(148) and Eq.~(156) of \cite{cannell}.
The expressions for $B^{(1)}, B^{(2)}$ can be found in Eqs.~(67) and Eq.~(148) of \cite{cannell}.
The terms $\propto \bm k'_x$ are also found in \cite{cannell}, when one identifies $\bm k'_x$ with the equivalent variable $\sin \theta_a$ there. 
Note $\delta n_z \propto \sin(q x)$  in \cite{cannell}.
Thus the coordinate shift  $x \rightarrow \pi /(2 q)$  is necessary to map their results to our convention $\delta n_z \propto \cos(q x)$ according to Eq.~(\ref{eq:ansatz_n}).
\end{thebibliography}
\end{document}